# Preamble-Based Channel Estimation for CP-OFDM and OFDM/OQAM Systems: A Comparative Study

Dimitris Katselis, Eleftherios Kofidis, Athanasios Rontogiannis, and Sergios Theodoridis

**Abstract**

In this paper, preamble-based least squares (LS) channel estimation in OFDM systems of the QAM and offset QAM (OQAM) types is considered, in both the frequency and the time domains. The construction of optimal (in the mean squared error (MSE) sense) preambles is investigated, for both the cases of full (all tones carrying pilot symbols) and sparse (a subset of pilot tones, surrounded by nulls or data) preambles. The two OFDM systems are compared for the same transmit power, which, for cyclic prefix (CP) based OFDM/QAM, also includes the power spent for CP transmission. OFDM/OQAM, with a sparse preamble consisting of equipowered and equispaced pilots embedded in zeros, turns out to perform at least as well as CP-OFDM. Simulations results are presented that verify the analysis.

## I. Introduction

Orthogonal frequency division multiplexing (OFDM) is currently enjoying popularity in both wired and wireless communication systems [2], mainly because of its immunity to multipath fading, which allows for a significant increase in the transmission rate [23]. Using the cyclic prefix (CP) as a guard interval, OFDM can "reform" a frequency selective channel into a set of parallel flat channels with independent noise disturbances. This greatly simplifies both the estimation of the channel as well as the recovery of the transmitted data at the receiver. However, these advantages come at the cost of an increased sensitivity to frequency offset and Doppler spread. This is due to the fact that, although the subcarrier functions are perfectly localized in time, they suffer from spectral leakage

D. Katselis and S. Theodoridis are with the Department of Informatics and Telecommunications, University of Athens, Panepistimioupolis, Ilissia, 157 84 Athens, Greece. E-mail: {dimitrik,stheodor}@di.uoa.gr.

E. Kofidis is with the Department of Statistics and Insurance Science, University of Piraeus, 80, Karaoli & Dimitriou str., 185 34 Piraeus, Greece. E-mail: kofidis@unipi.gr.

A. Rontogiannis is with the Institute for Space Applications and Remote Sensing, National Observatory of Athens, 152 36 P. Penteli, Greece. E-mail: tronto@space.noa.gr.

in the frequency domain. Moreover, the inclusion of the CP entails a loss in spectral efficiency, which, in practical systems, can become as high as 25% [2].

An alternative to CP-OFDM, that can mitigate these drawbacks, is provided by a filter bank-based variant employing offset quadrature amplitude modulation (OQAM), known as OFDM/OQAM [12]. This scheme builds upon a pulse shaping, which is achieved via (a) an IFFT/FFT-based efficient filter bank, and (b) staggered OQAM symbols; i.e., real symbols, at twice the symbol rate of OFDM/QAM, are loaded on the subcarriers [21]. This allows for the pulses to be well localized in both the time and the frequency domains. As a consequence, the system's robustness to frequency offsets and Doppler effects is increased [11] and at the same time an enhanced spectral containment, for bandwidth sensitive applications, is offered [1], [22]. Furthermore, although the two OFDM schemes can be seen to exhibit similar peak-to-average power ratio (PAPR) performances, the presence of spectral leakage in OFDM/QAM may, ultimately, generate higher peak power values [24]. Moreover, the use of a CP is not required in the OFDM/OQAM transmission, which may lead to even higher transmission rates [21].[1]

Since nothing is free in this world, the previously mentioned advantages of the OFDM/OQAM come at the cost of subcarrier functions being now orthogonal only in the real field, which means that there is always an *intrinsic* imaginary interference among (adjacent) subcarriers [9]. This makes the channel estimation task for OFDM/OQAM systems more challenging, compared to OFDM/QAM. OFDM/OQAM channel estimation has been recently studied for both preamble-based [13], [15] and scattered pilots-based [9], [14] training schemes.

The focus of this paper is on the channel estimation task based on a preamble consisting of pilot tones. The question of selecting these tones so as to minimize the channel estimation mean squared error (MSE), subject to a given training energy, is addressed. The cases of a *full* preamble, where *all* subcarriers carry pilots, and a *sparse* preamble, built upon isolated pilot tones embedded in nulls, are separately treated.[2] It is shown that an optimal sparse preamble for OFDM/OQAM can be constructed with $L_h$ equispaced and equipowered pilot tones, where $L_h$ denotes the channel length.

Related results have previously been derived for the case of CP-based OFDM/QAM (CP-OFDM) channel estimation. In [19], it is shown that uniform spacing is the best choice given that the pilot tones are equipowered.[3] Equispaced and equipowered pilot tones were shown in [3] to be the optimal CP-OFDM preamble for a given training energy that accounts only for the useful signal, excluding the CP. This paper also revisits the problem of optimally selecting the pilot tones in CP-OFDM, when the training energy constraint also includes the CP part. It is shown that, in this case, the pilots should also be *equal*. The effects of such a choice on the resulting PAPR are also discussed. For optimal CP-OFDM full preambles, it turns out that they can contain simply equipowered (not necessarily equal) symbols. A method of constructing such vectors is also developed. In OFDM/OQAM, all equal pilots result in optimal full preambles.

[1]Nevertheless, this advantage was partly given up in [16] and a CP-based OFDM/OQAM system was proposed for the sake of facilitating the data reception process.

[2]These pilot arrangements are also referred to as *block-type* and *comb-type*, respectively [5].

[3]This is no longer valid if there are suppressed (*virtual*) subcarriers. In such a case, the optimal placement is non-uniform [18].

In the sequel, the task of using extra (more than the minimum required $L_h$) pilot tones in a sparse preamble is considered and it is shown that no extra gain is provided. Furthermore, the case of including data symbols in addition to the pilots, in order to save bandwidth, is also considered and it is shown to result in a performance degradation for both OFDM systems. Full and sparse preambles are compared and turn out to ultimately yield the same estimation performance. The comparison of optimal sparse preambles for CP-OFDM and OFDM/OQAM turns out to be generally in favor of the latter. We present simulations results that confirm the theoretical analysis.

The rest of the paper is organized as follows: In Section II, we describe the discrete-time baseband equivalent model for the OFDM/OQAM and CP-OFDM systems. The way the various preambles are compared, in terms of MSE performance, is detailed in Section III. Necessary for the paper definitions and results are summarized in Section IV. Section V is devoted to the comparative study of the full versus the sparse preamble vectors, for both the OFDM/OQAM and CP-OFDM systems. The use of $P > L_h$ pilot tones in a sparse preamble is investigated in Section VI. Our results concerning the various scenarios of including data with the preamble are briefly presented in Section VII. In Section VIII, the MSE performances of the optimal sparse preambles associated with the two systems are compared. An error floor analysis concerning the OFDM/OQAM system is presented in Section IX. Simulations results are reported in Section X. Section XI concludes the paper.

*Notation.* Vectors and matrices are denoted by bold lowercase and uppercase letters, respectively. Superscripts $^T$ and $^H$ stand for transposition and conjugate transposition. The complex conjugate of a complex number $z$ is denoted by $z^*$. Also, $\jmath = \sqrt{-1}$. $\|\cdot\|$ is the Euclidean norm. For a matrix $\boldsymbol{A}$, $(\boldsymbol{A})_{i,j}$ denotes its $(i,j)$ entry. The expectation and matrix trace operators are denoted by $E(\cdot)$ and $\mathrm{tr}(\cdot)$, respectively. $\boldsymbol{I}_m$ denotes the $m$th-order identity matrix, while $\mathbf{0}_{m\times n}$ is the all zeros $m\times n$ matrix.

## II. System Models

In this section, basic definitions of the CP-OFDM and OFDM/OQAM system models are presented, along with some basic concepts that will be used in the sequel.

### A. CP-OFDM

Given $M$ subcarriers, the result of the OFDM modulation of a (complex) $M\times 1$ vector $\boldsymbol{x}$ is

$$\boldsymbol{s} = \frac{1}{\sqrt{M}}\boldsymbol{\mathcal{F}}^H\boldsymbol{x}$$

where $\boldsymbol{\mathcal{F}}$ is the $M\times M$ DFT matrix, with entries $(\boldsymbol{\mathcal{F}})_{i,j} = e^{-\jmath\frac{2\pi}{M}ij}$, $i,j = 0,1,\ldots,M-1$. Prior to transmission, a CP of length $\nu$ is prepended to the previous vector, to yield:

$$\boldsymbol{s}_{\mathrm{QAM}} = \begin{bmatrix} \mathbf{0}_{\nu\times(M-\nu)} \quad \boldsymbol{I}_\nu \\ \cdots\cdots\cdots\cdots\cdots\cdots \\ \boldsymbol{I}_M \end{bmatrix}\boldsymbol{s} \tag{1}$$

Assume that the CP length is chosen to be the smallest possible one, namely equal to the channel order: $\nu = L_h - 1$ [19]. Moreover, perfect timing and frequency synchronization is assumed. The channel impulse response (CIR), $\boldsymbol{h} = \left[ \begin{array}{cccc} h_0 & h_1 & \cdots & h_{L_h-1} \end{array} \right]^T$, is assumed to be constant over the duration of an OFDM symbol. The input to the OFDM demodulator, after the CP removal, can then be expressed as

$$\boldsymbol{r} = \boldsymbol{\mathcal{H}}\boldsymbol{s} + \boldsymbol{w},$$

where $\boldsymbol{\mathcal{H}}$ is the Toeplitz circulant matrix with the first row given by $\left[ \begin{array}{cccccc} h_0 & \mathbf{0}_{1\times(M-L_h)} & h_{L_h-1} & \cdots & h_2 & h_1 \end{array} \right]$ and $\boldsymbol{w}$ is the noise at the receiver front end and it is assumed to be white Gaussian with zero-mean and variance $\sigma^2$. The action of the DFT (FFT) then results in

$$\boldsymbol{y} = \frac{1}{\sqrt{M}}\boldsymbol{\mathcal{F}}\boldsymbol{r} = \text{diag}\left(H_0, H_1, \ldots, H_{M-1}\right)\boldsymbol{x} + \boldsymbol{\eta} \tag{2}$$

where $H_m = \sum_{l=0}^{L_h-1} h_l e^{-\jmath\frac{2\pi}{M}ml}$, $m = 0, 1, \ldots, M-1$ is the $M$-point channel frequency reponse (CFR) and $\boldsymbol{\eta} = \frac{1}{\sqrt{M}}\boldsymbol{\mathcal{F}}\boldsymbol{w}$ is the frequency domain noise, with the same statistics as $\boldsymbol{w}$. The CFR estimates, in the least squares (LS) sense, can then be computed as

$$\hat{H}_m = \frac{y_m}{x_m} = H_m + \frac{\eta_m}{x_m} \tag{3}$$

### B. OFDM/OQAM

The baseband discrete-time signal at time instant $l$, at the output of an OFDM/OQAM synthesis filter bank (SFB) is given by [21]:

$$s_{\text{OQAM}}(l) = \sum_{m=0}^{M-1}\sum_{n} a_{m,n} g_{m,n}(l) \tag{4}$$

where $a_{m,n}$ are real OQAM symbols, and

$$g_{m,n}(l) = g\left(l - n\frac{M}{2}\right) e^{\jmath\frac{2\pi}{M}m\left(l-\frac{L_g-1}{2}\right)} e^{\jmath\varphi_{m,n}},$$

with $g$ being the *real symmetric* prototype filter impulse response (assumed here of unit energy) of length $L_g$, $M$ being the *even* number of subcarriers, and $\varphi_{m,n} = \varphi_0 + \frac{\pi}{2}(m+n) \bmod \pi$, where $\varphi_0$ can be arbitrarily chosen[4] [21]. The filter $g$ is usually designed to have length $L_g = KM$, where $K$, the overlapping factor, takes on values in $1 \leq K \leq 5$ in practice. The double subscript $(\cdot)_{m,n}$ denotes the $(m,n)$-th time-frequency (TF) point. Thus, $m$ is the subcarrier index and $n$ the OQAM symbol time index.[5]

The pulse $g$ is designed so that the associated subcarrier functions $g_{m,n}$ are orthogonal in the real field, that is

$$\Re\left\{\sum_{l} g_{m,n}(l) g_{p,q}^{*}(l)\right\} = \delta_{m,p}\delta_{n,q}, \tag{5}$$

where $\delta_{i,j}$ is the Kronecker delta (i.e., $\delta_{i,j} = 1$ if $i = j$ and $0$ otherwise). This implies that even in the absence of channel distortion and noise, and with perfect time and frequency synchronization, there will be some intercarrier (and/or intersymbol) interference at the output of the analysis filter bank (AFB), which is purely imaginary, i.e.,

$$\sum_{l} g_{m,n}(l) g_{p,q}^{*}(l) = \jmath u_{m,n}^{p,q}, \tag{6}$$

[4] For example, in [21], $\varphi_{m,n}$ is defined as $(m+n)\frac{\pi}{2} - mn\pi$.

[5] The latter should not be confused with the sample time index $l$. In fact, the temporal distance between two successive symbol instants $n, n+1$ equals $M/2$ sample time instants.

and it is known as *intrinsic* interference [9]. Adopting the commonly used assumption that the channel is (approximately) frequency flat at each subcarrier and constant over the duration of the prototype filter [13], which is true for practical values of $L_h$ and $L_g$ and for well time-localized $g$'s, one can express the AFB output at the $p$th subcarrier and $q$th OFDM/OQAM symbol as:

$$y_{p,q} = H_{p,q}a_{p,q} + \jmath \underbrace{\underbrace{\sum_{m=0}^{M-1}\sum_{n}}_{(m,n)\neq(p,q)} H_{m,n}a_{m,n}u_{m,n}^{p,q}}_{I_{p,q}} + \eta_{p,q} \tag{7}$$

where $H_{p,q}$ is the CFR at that TF point, and $I_{p,q}$ and $\eta_{p,q}$ are the associated interference and noise components, respectively. One can easily see that $\eta_{p,q}$ is also Gaussian with zero mean and variance $\sigma^2$.

For pulses $g$ that are well localized in both time and frequency, the interference from TF points outside a neighborhood $\Omega_{p,q}$ around $(p,q)$ is negligible. If, moreover, the CFR is almost constant over this neighborhood, one can write (7) as

$$y_{p,q} \approx H_{p,q}c_{p,q} + \eta_{p,q} \tag{8}$$

where

$$c_{p,q} = a_{p,q} + \jmath \sum_{(m,n)\in\Omega_{p,q}} a_{m,n}u_{m,n}^{p,q} \tag{9}$$

When pilots are transmitted at $(p,q)$ and at points inside its neighborhood $\Omega_{p,q}$, the quantity in (9) can be approximately computed. This can then serve as a *pseudo-pilot* [13] to compute an estimate of the CFR at the corresponding TF point, as

$$\hat{H}_{p,q} = \frac{y_{p,q}}{c_{p,q}} \approx H_{p,q} + \frac{\eta_{p,q}}{c_{p,q}} \tag{10}$$

With a well time-frequency localized pulse, contributions to $I_{p,q}$ only come from the first-order neighborhood of $(p,q)$, namely $\Omega_{p,q} = \{(p\pm1, q\pm1), (p, q\pm1), (p\pm1, q)\}$. A special case is given by $\Omega_{p,q} = \Omega_{p,q}^1 = \{(p\pm1, q)\}$. This arises when we place three adjacent pilot tones at positions $(p-1,q), (p,q), (p+1,q)$ and zeros at the rest of the first-order neighborhood positions or when we place nonzero pilot tones at all positions in the preamble vector and zero vectors around it. If we abuse OQAM modulation (*only* in the preamble vector) by transmitting the complex symbols $a_{p,q}e^{\jmath\theta}$, $p = 0, 1, \ldots, M-1, \theta \in \{0, \pi, \pm\pi/2\}$, then, for an arbitrary $p$, the corresponding pseudo-pilot becomes:

$$c_{p,q} = a_{p,q} + \sum_{(m,n)\in\Omega_{p,q}^1} a_{m,n}u_{m,n}^{p,q}, \tag{11}$$

which is real. This is because by using the same phase factors $e^{\jmath\theta}$ in all the subcarriers we get:

$$\sum_{l} g_{m,n}(l)g_{p,q}^{*}(l) = u_{m,n}^{p,q}, \quad (m,n) \in \Omega_{p,q}^1 \tag{12}$$

If the first-order neighbors of $(p,q)$ carry unknown (data) symbols, one cannot approximate the imaginary interference in (9). However, by properly choosing *one* of the neighboring symbols, say at the point $(r,s)$, this interference can be forced to zero. Then the pseudo-pilot in (9) becomes real and equal to $a_{p,q}$. The pilot at $(r,s)$ is then known as a *help pilot* [9].

## III. A Fair Comparison Framework

For the CP-OFDM system, the preamble structure will consist of one complex vector symbol, as it is common in the literature [19]. Note that each complex CP-OFDM symbol is equivalent to two real vector symbols in the OFDM/OQAM system. We consider an equivalent preamble structure for the OFDM/OQAM system, which consists of one nonzero training vector followed by a zero vector symbol. The latter aims at protecting the nonzero part of the preamble from the intrinsic interference due to the data section of the frame [13]. Note that in wireless standards (e.g., WiMAX [2]), there are sufficiently long guard periods between the uplink and downlink subframes and between frames. Thus, there is no need to worry about intrinsic interference on the preamble vector from previous frames. Let us make this more clear:

*Definition 1:* Let a preamble structure consist of a number of training vector symbols with only one of them being nonzero. This nonzero vector will be called the *preamble vector*.

Let $T_1, T_2$ be the sampling periods at the SFB outputs for two OFDM systems. Assume that the minimum required number of SFB output samples to reconstruct the preamble vector at the receiver is $R_1, R_2$ for each system, respectively. Then the following quantity will be needed in making a fair comparison of the two preambles:

*Definition 2:* The training power ratio for the preamble structures of two systems is defined as:

$$\mathrm{TPR}^{\boldsymbol{p}_1,\boldsymbol{p}_2} = \frac{\frac{1}{R_1 T_1}\mathcal{E}^{\boldsymbol{p}_1}}{\frac{1}{R_2 T_2}\mathcal{E}^{\boldsymbol{p}_2}}$$

where $\mathcal{E}^{\boldsymbol{p}_1}, \mathcal{E}^{\boldsymbol{p}_2}$ are the energies of the corresponding preambles at the SFB outputs in the minimum sample numbers $R_1, R_2$ respectively.

Note that, if the two systems are of the same type, then $T_1 = T_2$ and the sampling periods can be omitted in the last definition. We can now clarify what will a fair comparison will be:

*Remark:* Assume that $\boldsymbol{p}_1, \boldsymbol{p}_2$ are different preamble structures in two systems. Then in order to guarantee a fair comparison between these two preamble structures, it is necessary that $\mathrm{TPR}^{\boldsymbol{p}_1,\boldsymbol{p}_2} = 1$. In other words, we will require that the systems under comparison spend the same power on the training data at the transmit antenna. If the training power ratio is not equal to one, we can scale the output of the SFB for the second preamble by $\sqrt{\mathrm{TPR}^{\boldsymbol{p}_1,\boldsymbol{p}_2}}$ to equalize the training powers for the two preamble structures.

## IV. Summary of Definitions and Useful Results

Throughout this paper, we make the assumption that $M/L_h$ is an integer number, with $M$ being (as usually in practice) a power of two. The channel length, $L_h$, will thus also be assumed to be a power of two.[6] Moreover, we will assume (as usual) that we have a *sample-spaced* channel [6] and that the nonzero part of the CIR is concentrated on its first $L_h$ taps.

*Definition 3:* By *sparse* preamble vector we will mean an $M \times 1$ training vector containing $L_h$ isolated pilots and zeros at the rest of its entries.

[6] It this is not the case, one may zero pad the CIR to $2^{\lceil \log_2 L_h \rceil}$ taps.

*Definition 4:* A preamble vector will be said to be *full* if it contains pilots at all of its entries.

*Definition 5:* A preamble vector with $L_h$ isolated pilots and data symbols at the rest of its entries will be called *sparse-data* preamble vector.

The following results will be useful in the sequel, hence we briefly summarize them here. Their proofs are in the appendices.

*Theorem 1:* For CP-OFDM, the sparse preamble that minimizes the MSE of the CFR estimates (3), subject to a constraint on the energy of *both* the useful part of the transmitted signal and the CP, consists of *equispaced* and *equal* pilot tones.

*Proof:* See Appendices I, II. ∎

*Remark.* Recall that the MSE-optimal sparse preamble for CP-OFDM is built with equispaced and *equipowered*, not necessarily equal pilots, if for the training energy the CP part is not included [19]. However, at least in theory this is not fair since the total amount of energy actually spent for training includes the transmission of the CP as well. According to Theorem 1, if the preamble optimization is to be performed in a fair way, then the pilots should necessarily be all equal. Of course, such a preamble would suffer from a high PAPR. In practice, this can be overcome by transmitting equipowered instead of equal tones, at the expense of a small (practically negligible) performance loss.

Although the above result holds true for the corresponding full preamble as well, we also prove that:

*Theorem 2:* There are full preambles for the CP-OFDM system that are MSE-optimal subject to a total energy budget both on the useful part of the transmitted signal and the CP, which result in equipowered but *unequal* pilot tones.

*Proof:* See Appendix III. ∎

This result indicates that we can construct optimal full preamble vectors that do not suffer from high PAPR values. The construction proposed in Appendix III is not unique and might be generalized. However, the proposed algorithm provides an infinite possible number of such full preamble vectors, verifying that a low PAPR full preamble vector construction is possible.

For the OFDM/OQAM system, the corresponding results are:

*Theorem 3:* For OFDM/OQAM, the sparse preamble that minimizes the MSE of the CFR estimates (10), subject to an energy constraint, is built with *equispaced* and *equipowered* pilot tones.

*Proof:* See Appendix IV. ∎

*Theorem 4:* Full OFDM/OQAM preambles with all *equal* pilots are locally MSE-optimal subject to a transmit energy constraint. Their global MSE-optimality is assured when the transmit energy constraint is translated at the input of the SFB.

*Proof:* See Appendix IV. ∎

It should be noted that, for the OFDM/OQAM system, the pilot symbols incorporate the corresponding phase factor $e^{\jmath\varphi_{m,n}}$. In view of the above results, preamble vectors containing equal symbols are the only or one of the optimal solutions. For the sake of analytical convenience, optimal preambles will henceforth be assumed to consist of all equal pilots.

## V. Full vs. Sparse Preamble

Let $\mathcal{I}_{L_h} = \left\{ i_0 + k\frac{M}{L_h} \middle| k = 0, 1, \ldots, L_h - 1 \right\}$, for a fixed preselected $i_0 \in \left\{0, 1, \ldots, \frac{M}{L_h} - 1\right\}$, be the set of indices of the nonzero pilot tones in the sparse preamble. We denote by $\mathcal{I}_{\overline{L_h}}$ the set $\{0, 1, \ldots, M-1\} \setminus \mathcal{I}_{L_h}$ of the remaining indices. Then, the full preamble vector can be written as:

$$\boldsymbol{x} = \boldsymbol{x}_{L_h} + \boldsymbol{x}_{\overline{L_h}} \tag{13}$$

where $\boldsymbol{x}_{L_h}$ is an $M \times 1$ vector with the $L_h$ nonzero pilots at the positions dictated by $\mathcal{I}_{L_h}$ and zeros elsewhere, while $\boldsymbol{x}_{\overline{L_h}}$ is the $M \times 1$ vector containing the rest of the pilot tones in the full preamble vector at the positions dictated by $\mathcal{I}_{\overline{L_h}}$ and zeros elsewhere. It is evident that $\boldsymbol{x}_{L_h}^H \boldsymbol{x}_{\overline{L_h}} = 0$ and $\|\boldsymbol{x}\|^2 = \|\boldsymbol{x}_{L_h}\|^2 + \|\boldsymbol{x}_{\overline{L_h}}\|^2$.

### A. OFDM/OQAM

The output of the SFB corresponding to the preamble section is then given by (cf. (4) with $n = 0$):

$$s_{\mathrm{OQAM}}(l) = \sum_{m=0}^{M-1} a_{m,0} g_{m,0}(l), \quad l = 0, 1, \ldots, L_g + \frac{M}{2} - 1$$

with the nonzero samples located at $k = 0, 1, \ldots, L_g - 1$. Incorporating the phase factors $e^{\jmath\varphi_{m,0}}$ into the real-valued symbols $a_{m,0}$, we obtain the complex training symbols $x_{m,0} = a_{m,0} e^{\jmath\varphi_{m,0}}$. Then imposing the restriction that all the training symbols are equal, the phase factors can also, without loss of generality, be considered all equal, say $e^{\jmath\varphi}$. Note that the requirement of equal symbols essentially leads to an abuse of the OQAM modulation, however this happens *only in the preamble section*. This has already been used in [15] in order to enhance the channel estimation performance. Using equal symbols of magnitude $\sqrt{\mathcal{E}_x^{\mathrm{OQAM}}}$, the last expression is then written as:

$$s_{\mathrm{OQAM}}(l) = e^{\jmath\varphi} \sqrt{\mathcal{E}_x^{\mathrm{OQAM}}} \left[ \sum_{m \in \mathcal{I}_{L_h}} g'_{m,0}(l) + \sum_{m \in \mathcal{I}_{\overline{L_h}}} g'_{m,0}(l) \right]$$

where $g'_{m,n}(l) = g\left(l - n\frac{M}{2}\right) e^{\jmath \frac{2\pi}{M} m \left(l - \frac{L_g - 1}{2}\right)}$. We may stack the *nonzero* output samples, corresponding to the preamble vector, to get $\boldsymbol{s}_{\mathrm{OQAM}} = \left[ \begin{array}{cccc} s_{\mathrm{OQAM}}(0) & s_{\mathrm{OQAM}}(1) & \cdots & s_{\mathrm{OQAM}}(L_g - 1) \end{array} \right]^T = \boldsymbol{s}_{\mathrm{OQAM}}^{L_h} + \boldsymbol{s}_{\mathrm{OQAM}}^{\overline{L_h}}$, where $\boldsymbol{s}_{\mathrm{OQAM}}^{L_h} = e^{\jmath\varphi} \sqrt{\mathcal{E}_x^{\mathrm{OQAM}}} \left[ \begin{array}{cccc} \sum_{m \in \mathcal{I}_{L_h}} g'_{m,0}(0) & \sum_{m \in \mathcal{I}_{L_h}} g'_{m,0}(1) & \cdots & \sum_{m \in \mathcal{I}_{L_h}} g'_{m,0}(L_g - 1) \end{array} \right]^T$ and similarly for $\boldsymbol{s}_{\mathrm{OQAM}}^{\overline{L_h}}$.

The energy of the SFB output corresponding to the full preamble can be expressed as:

$$\mathcal{E}_{\mathrm{OQAM}}^{\mathrm{f}} = \|\boldsymbol{s}_{\mathrm{OQAM}}\|^2 = M(1 + 2\beta)\mathcal{E}_x^{\mathrm{OQAM}} \tag{14}$$

while the energy of the sparse preamble as:

$$\mathcal{E}_{\mathrm{OQAM}}^{\mathrm{s}} = \|\boldsymbol{s}_{\mathrm{OQAM}}^{L_h}\|^2 = L_h \mathcal{E}_x^{\mathrm{OQAM}}, \tag{15}$$

as it easily follows from the analysis in Appendix IV. A proper analytical description for the quantity $\beta$, related to the intrinsic interference from adjacent subcarriers, will be given later on. The power ratio for the full and sparse preamble structures of the OFDM/OQAM system is thus given by

$$\mathrm{TPR}_{\mathrm{OQAM}}^{\mathrm{f,s}} = \frac{M(1+2\beta)}{L_h},$$

where the superscript $\mathrm{f,s}$ stands for the words *f*ull over *s*parse.

Next, we will evaluate the channel estimation performance of the two preambles in the frequency domain. Let us first consider the sparse preamble. The outputs of the AFB at the pilot positions are given by:

$$y_{m,0} = H_{m,0}a_{m,0} + \eta_{m,0}, \quad m \in \mathcal{I}_{L_h}$$

For the sparse preamble to be meaningful, there should hold that $M/L_h \geq 2$. If the pulse is well-localized in frequency, we can assume that the received training samples are uncorrelated, since none of them belongs to the first-order neighborhood of the rest. Therefore, the maximum likelihood (ML) estimator coincides with the LS estimator, and the CFR can be estimated as:

$$\hat{H}_{m,0} = \frac{y_{m,0}}{a_{m,0}} = H_{m,0} + \frac{\eta_{m,0}}{a_{m,0}}, \quad m \in \mathcal{I}_{L_h} \tag{16}$$

Stacking these estimates in the vector $\hat{\boldsymbol{H}}_{L_h}$ and recalling our assumption of a sample-spaced channel, we can obtain an estimate of the CIR as:

$$\hat{\boldsymbol{h}} = \boldsymbol{F}_{L_h \times L_h}^{-1} \hat{\boldsymbol{H}}_{L_h} = \boldsymbol{h} + \boldsymbol{F}_{L_h \times L_h}^{-1} \boldsymbol{\eta}'_{L_h},$$

where $\boldsymbol{\eta}'_{L_h} = \begin{bmatrix} \frac{\eta_{i_0,0}}{a_{i_0,0}} & \frac{\eta_{i_0+M/L_h,0}}{a_{i_0+M/L_h,0}} & \cdots & \frac{\eta_{i_0+(L_h-1)M/L_h,0}}{a_{i_0+(L_h-1)M/L_h,0}} \end{bmatrix}^T$, and $\boldsymbol{F}_{L_h \times L_h}$ is the $L_h \times L_h$ submatrix of the $M \times M$ DFT matrix $\boldsymbol{\mathcal{F}}$ consisting of its first $L_h$ columns and its rows corresponding to the indices in $\mathcal{I}_{L_h}$. The CFR at all $M$ subcarriers can then be recovered by:

$$\hat{\boldsymbol{H}} = \boldsymbol{F}_{M \times L_h} \hat{\boldsymbol{h}} = \boldsymbol{H} + \boldsymbol{F}_{M \times L_h} \boldsymbol{F}_{L_h \times L_h}^{-1} \boldsymbol{\eta}'_{L_h}$$

where $\boldsymbol{H} = \begin{bmatrix} H_{0,0} & H_{1,0} & \cdots & H_{M-1,0} \end{bmatrix}^T$, and $\boldsymbol{F}_{M \times L_h}$ denotes the $M \times L_h$ submatrix of $\boldsymbol{\mathcal{F}}$, consisting of its first $L_h$ columns. Denoting by $\boldsymbol{\mathcal{C}}_{L_h}$ the covariance matrix of $\boldsymbol{\eta}'_{L_h}$, the MSE of the latter estimator is given by:

$$\mathrm{MSE}_{\mathrm{OQAM}}^{\mathrm{s}} = E\left[\left\|\hat{\boldsymbol{H}} - \boldsymbol{H}\right\|^2\right] = \mathrm{tr}\left(\boldsymbol{F}_{M \times L_h} \boldsymbol{F}_{L_h \times L_h}^{-1} \boldsymbol{\mathcal{C}}_{L_h} \boldsymbol{F}_{L_h \times L_h}^{-H} \boldsymbol{F}_{M \times L_h}^{H}\right)$$

By our assumptions, $\boldsymbol{\mathcal{C}}_{L_h} = \frac{\sigma^2}{\mathcal{E}_x^{\mathrm{OQAM}}} \boldsymbol{I}_{L_h}$. Additionally, it is known ([19]) that $\boldsymbol{F}_{L_h \times L_h} \boldsymbol{F}_{L_h \times L_h}^{H} = \boldsymbol{F}_{L_h \times L_h}^{H} \boldsymbol{F}_{L_h \times L_h}$ $= L_h \boldsymbol{I}_{L_h}$. Also, using the fact that $\boldsymbol{F}_{M \times L_h}^{H} \boldsymbol{F}_{M \times L_h} = M \boldsymbol{I}_{L_h}$, we finally obtain:

$$\mathrm{MSE}_{\mathrm{OQAM}}^{\mathrm{s}} = \frac{M\sigma^2}{\mathcal{E}_x^{\mathrm{OQAM}}} \tag{17}$$

Special care is needed in the full preamble case. For the assumed construction of the preamble and with a well frequency-localized pulse, we can express each output sample of the AFB by:

$$y_{m,0} = H_{m,0}\left(a_{m,0} + \sum_{l \in \{-1,+1\}} a_{m+l,0} u_{m+l,0}^{m,0}\right) + \eta_{m,0} = H_{m,0} c_{m,0} + \eta_{m,0}, \quad m = 0, 1, \ldots, M-1$$

where the same arguments as in the derivation of eq. (11) have been used. In the last equation, for $m = 0$ the value $m-1$ corresponds to the $(M-1)$th subcarrier and for $m = M-1$ the value $m+1$ corresponds to the 0th subcarrier, due to spectrum periodicity in discrete-time. Furthermore, it is easy to show that for any real and symmetric prototype function $g$, we have $u_{m+1,0}^{m,0} = u_{m-1,0}^{m,0} = \beta$, $m = 0, 1, \ldots, M-1$, or $u_{m+1,0}^{m,0} = u_{m-1,0}^{m,0} = \beta$, $m = 1, 2, \ldots, M-2$, and $u_{1,0}^{0,0} = -u_{M-1,0}^{0,0}$, $u_{M-2,0}^{M-1,0} = -u_{0,0}^{M-1,0}$, depending on the value of $L_g$ in the factor $e^{-\jmath \frac{2\pi}{M} m \frac{L_g-1}{2}}$ (needed for causality purposes [21]) and due to its dependence on $m$. Let us consider the first case, although handling the second case is equally straightforward. In the light of these facts, we can see that:

$$c_{m,0} = \sqrt{\mathcal{E}_x^{\mathrm{OQAM}}}(1+2\beta), \quad m = 0, 1, \ldots, M-1$$

The MSE expression for the full preamble is thus easily seen to be given by:

$$\mathrm{MSE}_{\mathrm{OQAM}}^{\mathrm{f}} = \frac{M\sigma^2}{\mathcal{E}_x^{\mathrm{OQAM}}(1+2\beta)^2} \tag{18}$$

To make the comparison between the two preambles fair, we have first to equalize the powers at the outputs of the SFB's. Scaling the output of the SFB for the sparse preamble by $\sqrt{\mathrm{TPR}_{\mathrm{OQAM}}^{\mathrm{f,s}}}$, we achieve this goal. The previous analysis holds as is, with the only difference that the MSE for the sparse preamble is now given by:

$$\mathrm{MSE}_{\mathrm{OQAM}}^{\mathrm{s}} = \frac{M\sigma^2}{\mathcal{E}_x^{\mathrm{OQAM}}\mathrm{TPR}_{\mathrm{OQAM}}^{\mathrm{f,s}}} = \frac{L_h\sigma^2}{\mathcal{E}_x^{\mathrm{OQAM}}(1+2\beta)} \tag{19}$$

and the ratio of the two MSE's becomes:

$$\frac{\mathrm{MSE}_{\mathrm{OQAM}}^{\mathrm{s}}}{\mathrm{MSE}_{\mathrm{OQAM}}^{\mathrm{f}}} = (1+2\beta)\frac{L_h}{M}$$

Hence, in a dB scale, the sparse preamble is $10\log_{10}\{M/[L_h(1+2\beta)]\}$ better than the full preamble.

### *B. CP-OFDM*

By (2), and due to the complex field orthogonality of the DFT, the ML estimates of the CFR will again coincide with the LS estimates and will be as in (16). The analysis performed for the OFDM/OQAM sparse preamble applies also in the CP-OFDM sparse preamble case. Assuming that we transmit pilots of modulus $\sqrt{\mathcal{E}_x^{\mathrm{QAM}}}$, the MSE expression for the sparse preamble will be:

$$\mathrm{MSE}_{\mathrm{QAM}}^{\mathrm{s}} = \frac{M\sigma^2}{\mathcal{E}_x^{\mathrm{QAM}}},$$

which coincides with the MSE expression for the full preamble. Once more, to make a fair comparison, we have to evaluate the power ratio in this case. Using (1) and (13), the energy transmitted with the full preamble vector is

$$\mathcal{E}_{\mathrm{QAM}}^{\mathrm{f}} = \|\boldsymbol{s}_{\mathrm{QAM}}\|^2 = \|\boldsymbol{x}\|^2 + \underbrace{\frac{1}{M}\boldsymbol{x}^H \boldsymbol{F}_{M\times\nu}\boldsymbol{F}_{M\times\nu}^H\boldsymbol{x}}_{\text{CP energy}} \tag{20}$$

where $\boldsymbol{F}_{M\times\nu}$ is the $M \times \nu$ matrix consisting of the last $\nu$ columns of $\boldsymbol{\mathcal{F}}$. Clearly, $\|\boldsymbol{x}\|^2 = M\mathcal{E}_x^{\mathrm{QAM}}$. To evaluate the CP energy, we will need the following:

*Lemma 1:* For the matrix $\boldsymbol{F}_{M\times\nu}$, the following relationships hold:

$$\sum_i (\boldsymbol{F}_{M\times\nu}\boldsymbol{F}^H_{M\times\nu})_{i,i} = M\nu$$

and

$$\sum_{i,j} (\boldsymbol{F}_{M\times\nu}\boldsymbol{F}^H_{M\times\nu})_{i,j} = 0$$

*Proof:* For the first relationship, we have:

$$\sum_i (\boldsymbol{F}_{M\times\nu}\boldsymbol{F}^H_{M\times\nu})_{i,i} = \operatorname{tr}\left(\boldsymbol{F}_{M\times\nu}\boldsymbol{F}^H_{M\times\nu}\right) = \operatorname{tr}\left(\boldsymbol{F}^H_{M\times\nu}\boldsymbol{F}_{M\times\nu}\right) = M\nu$$

For the second one:

$$\sum_{i,j} (\boldsymbol{F}_{M\times\nu}\boldsymbol{F}^H_{M\times\nu})_{i,j} = \mathbf{1}^H_M \boldsymbol{F}_{M\times\nu}\boldsymbol{F}^H_{M\times\nu}\mathbf{1}_M = \|\boldsymbol{F}^H_{M\times\nu}\mathbf{1}_M\|^2 = 0$$

where $\mathbf{1}_M$ is the $M\times 1$ all ones vector. The last equation holds because $(1/M)\boldsymbol{F}^H_{M\times\nu}\mathbf{1}_M$ represents the last $\nu$ $M$-point IDFT coefficients of the rectangular pulse, which are all zero. ■

Consequently, since $\boldsymbol{x}$ has been assumed to be of the form $x\mathbf{1}_M$, with $|x| = \sqrt{\mathcal{E}_x^{\mathrm{QAM}}}$, the CP energy in (20) is *zero* and hence:

$$\mathcal{E}^{\mathrm{f}}_{\mathrm{QAM}} = M\mathcal{E}_x^{\mathrm{QAM}} \tag{21}$$

For the sparse preamble:

$$\mathcal{E}^{\mathrm{s}}_{\mathrm{QAM}} = \|\boldsymbol{x}_{L_h}\|^2 + \underbrace{\frac{1}{M}\boldsymbol{x}^H_{L_h}\boldsymbol{F}_{M\times\nu}\boldsymbol{F}^H_{M\times\nu}\boldsymbol{x}_{L_h}}_{\text{CP energy}} \tag{22}$$

Obviously,

$$\|\boldsymbol{x}_{L_h}\|^2 = L_h\mathcal{E}_x^{\mathrm{QAM}} \tag{23}$$

and for the CP part:

$$\begin{aligned}
\frac{1}{M}\boldsymbol{x}^H_{L_h}\boldsymbol{F}_{M\times\nu}\boldsymbol{F}^H_{M\times\nu}\boldsymbol{x}_{L_h} &= \frac{1}{M}\mathcal{E}_x^{\mathrm{QAM}}\sum_{l=1}^{\nu}\left|\sum_{m\in\mathcal{I}_{L_h}} e^{\jmath\frac{2\pi}{M}m(M-1-\nu+l)}\right|^2 \\
&= \frac{1}{M}\mathcal{E}_x^{\mathrm{QAM}}\sum_{l=1}^{\nu}\left|\sum_{m=0}^{L_h-1} e^{\jmath\frac{2\pi}{L_h}m(M-1-\nu+l)}\right|^2 = 0,
\end{aligned} \tag{24}$$

where we have used the fact that $\sum_{m=0}^{L_h-1} e^{\jmath\frac{2\pi}{L_h}m(M-1-\nu+l)}$ is zero for any value of $M-1-\nu+l$ that is not an integer multiple of $L_h$. Since $M-L_h+1 \leq M-1-\nu+l \leq M-1$, no such multiple exists.

Thus, the training power ratio for CP-OFDM is:

$$\mathrm{TPR}^{\mathrm{f,s}}_{\mathrm{QAM}} = \frac{\mathcal{E}^{\mathrm{f}}_{\mathrm{QAM}}}{\mathcal{E}^{\mathrm{s}}_{\mathrm{QAM}}} = \frac{M}{L_h} \tag{25}$$

Scaling by $\sqrt{\mathrm{TPR}^{\mathrm{f,s}}_{\mathrm{QAM}}}$ the output of the SFB for the sparse preamble, the associated MSE changes to:

$$\mathrm{MSE}^{\mathrm{s}}_{\mathrm{QAM}} = \frac{M\sigma^2}{\mathrm{TPR}^{\mathrm{f,s}}_{\mathrm{QAM}}\mathcal{E}_x^{\mathrm{QAM}}} = \frac{L_h\sigma^2}{\mathcal{E}_x^{\mathrm{QAM}}}$$

and, finally, the ratio of the MSE's for the two preambles is given by:

$$\frac{\mathrm{MSE}^{\mathrm{s}}_{\mathrm{QAM}}}{\mathrm{MSE}^{\mathrm{f}}_{\mathrm{QAM}}} = \frac{L_h}{M} \tag{26}$$

The last equation shows that the sparse preamble has a $10\log_{10}(M/L_h)$ dB better MSE performance than the full preamble. However, as it is shown below, this performance difference can be eliminated if the fact that the CIR is concentrated on $L_h$ taps is exploited in the estimation procedure.

### *C. A Certain Processing to Equalize the Performances*

Consider the previous analysis for the sparse and the full CP-OFDM preamble vectors. The estimates provided by the full-preamble vector can be viewed as being of the form:

$$\hat{\boldsymbol{H}} = \boldsymbol{H} + \sqrt{\frac{\sigma^2}{\mathcal{E}_x^{\mathrm{QAM}}}}\boldsymbol{\epsilon}$$

where $\boldsymbol{\epsilon}$ is an $M \times 1$ error vector, of zero mean and covariance $\boldsymbol{I}_M$. Consider the MSE for this preamble:

$$\mathrm{MSE}^{\mathrm{f}}_{\mathrm{QAM}} = E\left[\left\|\hat{\boldsymbol{H}} - \boldsymbol{H}\right\|^2\right] = \frac{\sigma^2}{\mathcal{E}_x^{\mathrm{QAM}}} E\left[\|\boldsymbol{\epsilon}\|^2\right] = \frac{\sigma^2}{\mathcal{E}_x^{\mathrm{QAM}}} M$$

To convert this estimate to the time domain, we apply to $\hat{\boldsymbol{H}}$ the transformation $\left(\boldsymbol{F}^H_{M\times L_h}\boldsymbol{F}_{M\times L_h}\right)^{-1}\boldsymbol{F}^H_{M\times L_h}$. If we want to bring it back to the frequency domain, we have to apply to the obtained $\hat{\boldsymbol{h}}$ the transformation $\boldsymbol{F}_{M\times L_h}$. This amounts to applying the transformation $\boldsymbol{F}_{M\times L_h}\left(\boldsymbol{F}^H_{M\times L_h}\boldsymbol{F}_{M\times L_h}\right)^{-1}\boldsymbol{F}^H_{M\times L_h}$ to the originally computed $\hat{\boldsymbol{H}}$. The MSE is now given by:

$$\mathrm{MSE}^{\mathrm{f}}_{\mathrm{QAM}} = \frac{\sigma^2}{\mathcal{E}_x^{\mathrm{QAM}}} E\left[\left\|\boldsymbol{F}_{M\times L_h}\left(\boldsymbol{F}^H_{M\times L_h}\boldsymbol{F}_{M\times L_h}\right)^{-1}\boldsymbol{F}^H_{M\times L_h}\boldsymbol{\epsilon}\right\|^2\right] = \frac{\sigma^2}{\mathcal{E}_x^{\mathrm{QAM}}} L_h$$

and therefore (26) now simplifies to:

$$\mathrm{MSE}^{\mathrm{f}}_{\mathrm{QAM}} = \mathrm{MSE}^{\mathrm{s}}_{\mathrm{QAM}} \tag{27}$$

Thus, this kind of processing thus leads to scaling the ratio of the MSE's by $M/L_h$. In Appendix V, we show that the corresponding effect for OFDM/OQAM is approximately equivalent to scaling the full preamble MSE by $L_h(1+2\beta)/M$, for practical values of $M, L_h$. Thus, applying this processing in the full preamble-based estimates in the OFDM/OQAM system, we again obtain:

$$\mathrm{MSE}^{\mathrm{s}}_{\mathrm{OQAM}} \approx \mathrm{MSE}^{\mathrm{f}}_{\mathrm{OQAM}}$$

*Remarks:*

1) $\boldsymbol{F}_{M\times L_h}\left(\boldsymbol{F}^H_{M\times L_h}\boldsymbol{F}_{M\times L_h}\right)^{-1}\boldsymbol{F}^H_{M\times L_h}$ projects $\hat{\boldsymbol{H}}$ onto the space of $M$-point CFR's with CIR's of length $L_h$. This has the effect of suppressing the estimation noise in the impulse response tail. In fact, the previous processing can be seen to be equivalent with constrained LS [7].
2) The previous analysis only holds for sample-spaced channels [20].
3) The MSE equivalence of full and sparse preambles is valid only for *optimal* preambles.

## VI. DO WE NEED MORE THAN $L_h$ PILOT TONES?

Let us now consider using a sparse preamble with $P > L_h$ equispaced and equal pilots to estimate the channel and compare its performance with that of the sparse preamble containing only $L_h$ pilots as before. For the OFDM/OQAM system, this preamble is still a sparse preamble so the optimality of equispaced and equipowered symbols holds as well. For the OFDM/QAM system, if the CP length is $P-1$, the sparse preamble with equispaced and equal pilot tones is the only optimal solution (see App. I,II). If the CP length is $L_h - 1$, the sparse preamble with equal and equispaced pilot tones is the optimal one as it is verified by the following analysis. The basic assumption is that $M/P$ is an integer. Additionally, $P < M$, since otherwise we end up with a full preamble. Based on our assumptions on $M, L_h$, it is easy to see that $P/L_h$ is an even integer. Also, the placement of the $P$ pilots follows the optimal rule, i.e., the nonzero subcarriers belong to any one of the sets $\left\{i_0, i_0 + \frac{M}{P}, \ldots, i_0 + (P-1)\frac{M}{P}\right\}$, $i_0 = 1, 2, \ldots, \frac{M}{P} - 1$, denoted by $\mathcal{I}_P$.

### A. OFDM/OQAM

One can easily verify (cf. (15)) that the training power ratio for $P$ pilots over $L_h$ pilots is:

$$\mathrm{TPR}_{\mathrm{OQAM}}^{\mathrm{P,Ls}} = \frac{\mathcal{E}_x^{\mathrm{OQAM}} \sum_{l=0}^{L_g-1} g^2(l) \left| \sum_{m=0}^{P-1} e^{\jmath \frac{2\pi}{P} m \left(l - \frac{L_g-1}{2}\right)} \right|^2}{\mathcal{E}_x^{\mathrm{OQAM}} \sum_{l=0}^{L_g-1} g^2(l) \left| \sum_{m=0}^{L_h-1} e^{\jmath \frac{2\pi}{L_h} m \left(l - \frac{L_g-1}{2}\right)} \right|^2} = \frac{P}{L_h} \tag{28}$$

Since $M/P \geq 2$, the spacing of the pilots in the preamble vector is larger than the size of the first-order neighborhood in the frequency direction. Relying again on the good frequency localization of the pulses, we can assume that there is almost no interference among the pilot symbols. To convert the $P$ CFR estimates to the time domain, we use in this case the transformation $\left(\boldsymbol{F}_{P\times L_h}^H \boldsymbol{F}_{P\times L_h}\right)^{-1} \boldsymbol{F}_{P\times L_h}^H$, where $\boldsymbol{F}_{P\times L_h}$ is the $P \times L_h$ submatrix of $\boldsymbol{\mathcal{F}}$ consisting of its first $L_h$ columns and its rows corresponding to the indices in $\mathcal{I}_P$. Then, the MSE for the sparse preamble with $P$ pilots is given by:

$$\mathrm{MSE}_{\mathrm{OQAM}}^{\mathrm{Ps}} = \mathrm{tr}\left[\boldsymbol{F}_{M\times L_h} \left(\boldsymbol{F}_{P\times L_h}^H \boldsymbol{F}_{P\times L_h}\right)^{-1} \boldsymbol{F}_{P\times L_h}^H \boldsymbol{\mathcal{C}}_P \boldsymbol{F}_{P\times L_h} \left(\boldsymbol{F}_{P\times L_h}^H \boldsymbol{F}_{P\times L_h}\right)^{-1} \boldsymbol{F}_{M\times L_h}^H\right]$$

where $\boldsymbol{\mathcal{C}}_P$ is the analogue of $\boldsymbol{\mathcal{C}}_{L_h}$ in this case. In view of the equal spacing of the $P$ pilots, we have $\boldsymbol{F}_{P\times L_h}^H \boldsymbol{F}_{P\times L_h} = P\boldsymbol{I}_{L_h}$, and, moreover, $\boldsymbol{\mathcal{C}}_P = \frac{\sigma^2}{\mathcal{E}_x^{\mathrm{OQAM}}} \boldsymbol{I}_P$. Using these results in the last expression, we obtain:

$$\mathrm{MSE}_{\mathrm{OQAM}}^{\mathrm{Ps}} = \frac{M\sigma^2}{\mathcal{E}_x^{\mathrm{OQAM}}} \frac{L_h}{P} \tag{29}$$

For the sparse preamble with $L_h$ pilots, (17) holds and hence, after the power equalization dictated by (28),

$$\mathrm{MSE}_{\mathrm{OQAM}}^{\mathrm{Ps}} = \mathrm{MSE}_{\mathrm{OQAM}}^{\mathrm{s}} \tag{30}$$

### B. CP-OFDM

For CP-OFDM, the analysis is similar. The MSE's are then given by:

$$\mathrm{MSE}_{\mathrm{QAM}}^{\mathrm{Ps}} = \frac{M\sigma^2}{\mathcal{E}_x^{\mathrm{QAM}}} \frac{L_h}{P}$$

and

$$\mathrm{MSE}_{\mathrm{QAM}}^{\mathrm{s}} = \frac{M\sigma^2}{\mathrm{TPR}_{\mathrm{QAM}}^{\mathrm{P,Ls}} \mathcal{E}_x^{\mathrm{QAM}}}$$

For the sparse preamble with $P$ pilots, the energy at the output of the SFB is given by $\mathcal{E}_{\mathrm{QAM}}^{\mathrm{Ps}} = \|\boldsymbol{x}_P\|^2 + \frac{1}{M}\boldsymbol{x}_P^H \boldsymbol{F}_{M\times\nu}\boldsymbol{F}_{M\times\nu}^H \boldsymbol{x}_P$, where $\boldsymbol{x}_P$ is the preamble vector containing $P$ equal pilots at positions dictated by $\mathcal{I}_P$ and zeros elsewhere. It is easy to see that $\mathcal{E}_{\mathrm{QAM}}^{\mathrm{Ps}} = P\mathcal{E}_x^{\mathrm{QAM}}$ since, with a CP of length $\nu = L_h - 1$,

$$\boldsymbol{x}_P^H \boldsymbol{F}_{M\times\nu}\boldsymbol{F}_{M\times\nu}^H \boldsymbol{x}_P = \mathcal{E}_x^{\mathrm{QAM}} \sum_{l=1}^{\nu} \left| \sum_{m\in\mathcal{I}_P} e^{\jmath\frac{2\pi}{M}m(M-1-\nu+l)} \right|^2 = \mathcal{E}_x^{\mathrm{QAM}} \sum_{l=1}^{\nu} \left| \sum_{m=0}^{P-1} e^{\jmath\frac{2\pi}{P}m(M-1-\nu+l)} \right|^2 = 0$$

as before. The training power ratio is easily shown to be $\mathrm{TPR}_{\mathrm{QAM}}^{\mathrm{P,Ls}} = P/L_h$ in this case as well, and therefore:

$$\mathrm{MSE}_{\mathrm{QAM}}^{\mathrm{Ps}} = \mathrm{MSE}_{\mathrm{QAM}}^{\mathrm{s}} \tag{31}$$

*Remark.* It thus turns out that using more pilot tones than suggested by the channel length would not result in any performance gain. Observe that this MSE equivalence again only holds for optimal (with equispaced and equal/equipowered pilot tones) preambles.

## VII. Including Data in the Preamble

What if the inactive tones in a sparse preamble are employed to carry data symbols? That would help saving part of the bandwidth consumed for training. In such a context, and in order to make a fair comparison between the preambles, the data power will not be considered as part of the training energy. This is because the data transmission is a benefit of the mixed (sparse-data) preamble. With this consideration, the implications of using such a preamble in each OFDM system can be easily explored using the previous analysis. We summarize some main results for the case when the data symbols have the same modulus with the training symbols. The case of a different pilot-to-data power ratio can be similarly handled.

### A. OFDM/OQAM

For this system, we consider two different preamble constructions, keeping the definition of the sparse preamble as we have done so far:

*1) Preamble with a nonzero vector followed by a zero side vector:* For this preamble, we have to consider two cases:

Scenario 1: The preamble vector contains data at all positions dictated by $\mathcal{I}_{\overline{L_h}}$. Since the data power is not taken into account in the equalization of the powers at the SFB outputs, the sparse-data and the sparse preambles result in the same training power at the output of the SFB. However, it can be easily proved for the resulting MSE that:

$$\mathrm{MSE}_{\mathrm{OQAM}}^{\mathrm{sd}} = \frac{M\sigma^2}{\mathcal{E}_x^{\mathrm{OQAM}}} + \frac{M}{L_h}\beta^2 \sum_{m\in\mathcal{I}_{L_h}} |H_{m,0}|^2 > \frac{M\sigma^2}{\mathcal{E}_x^{\mathrm{OQAM}}} = \mathrm{MSE}_{\mathrm{OQAM}}^{\mathrm{s}}$$

The last formula for $\mathrm{MSE}^{\mathrm{sd}}_{\mathrm{OQAM}}$ also implies that as $\sigma^2 \longrightarrow 0$ (or $\mathrm{SNR} \longrightarrow \infty$), the sparse data estimate will present an *error floor*, i.e.,

$$\mathrm{MSE}^{\mathrm{sd}}_{\mathrm{OQAM}} \underset{\sigma^2\to 0}{\longrightarrow} \frac{M}{L_h}\beta^2 \sum_{m\in\mathcal{I}_{L_h}} |H_{m,0}|^2$$

Scenario 2: We have data at all positions dictated by $\mathcal{I}_{\overline{L_h}}$, except for the positions around the pilot tones. In this case, the implicit assumption is that $M/L_h \geq 4$. Then, according to our assumptions, the intrinsic interference term in the previous scenario disappears and $\mathrm{MSE}^{\mathrm{sd}}_{\mathrm{OQAM}} = \mathrm{MSE}^{\mathrm{s}}_{\mathrm{OQAM}}$.

*2) Using a nonzero side vector containing data and help pilots:* In this preamble structure, the second column contains data at all positions except for the subcarriers in $\mathcal{I}_{L_h}$, which are loaded with the help pilots. Placing the help pilots at the same positions with the pilots is justified by PAPR considerations. In fact, it can be easily shown that, among the first-order neighbors of a TF point, the strongest intrinsic interference comes from those points corresponding to the same frequency. Specifically,

$$|u^{m,n}_{m,n\pm 1}| > |u^{m,n}_{m\pm 1,n}| > |u^{m,n}_{m\pm 1,n\pm 1}|$$

Consequently, placing the help pilots at the aforementioned positions leads to help pilots with smaller modulus, thus reducing the PAPR. Note that we can place the help pilots at the corresponding positions of the first column of the data section.

The above preamble structure aims at maximally exploiting the OFDM/OQAM system, by placing data at all available positions. We can then consider three different scenarios:

Scenario 1: The pilots are placed at the first column. The side column and the first column contain data at all positions except for the positions that belong to the first-order TF neighborhoods of each pilot tone. In this case, there is no need to use help pilots. Obviously, the sparse preamble and this preamble use the same training power. Since they will both use the same processing to get the channel estimates, they will lead to the same MSE.

Scenario 2: The first column contains the pilots and data at all other positions except for the positions adjacent to the pilots. The side column contains data and the help pilots at the aforementioned positions. Caution is then needed in the time durations we need to observe each preamble to collect the training energy. For the sparse preamble, this is $L_g$ samples, while for the sparse-data preamble it is $L_g + M/2$ samples. After some algebra, it can then be shown that the corresponding power ratio is:

$$\mathrm{TPR}^{\mathrm{sd,s}}_{\mathrm{OQAM}} = \frac{\frac{1}{L_g+\frac{M}{2}} L_h \mathcal{E}^{\mathrm{OQAM}}_x (1+\zeta)}{\frac{1}{L_g} L_h \mathcal{E}^{\mathrm{OQAM}}_x} = \frac{L_g(1+\zeta)}{L_g + \frac{M}{2}}$$

where

$$\zeta = \frac{\sum_{l\in\{\pm 1\}} \left(u^{m+l,1}_{m,0}\right)^2}{\left(u^{m,1}_{m,0}\right)^2} > 0, \quad \forall m \in \mathcal{I}_{L_h}$$

For $\zeta > M/(2L_g)$, we have $\mathrm{TPR}^{\mathrm{sd,s}}_{\mathrm{OQAM}} > 1$ and hence $\mathrm{MSE}^{\mathrm{s}}_{\mathrm{OQAM}} < \mathrm{MSE}^{\mathrm{sd}}_{\mathrm{OQAM}}$. Generally, the last inequalities hold, especially as $L_g$ increases for better TF localization of the prototype function.

Scenario 3: This case is similar to Scenario 2, where now we also place data at all positions adjacent to the pilot tones in the first column. The power of the help pilots will be even larger than that in Scenario 2, and therefore $\mathrm{MSE}^{\mathrm{s}}_{\mathrm{OQAM}} < \mathrm{MSE}^{\mathrm{sd}}_{\mathrm{OQAM}}$ in this case as well.

### B. CP-OFDM

In this context, $\boldsymbol{x}_{\overline{L_h}}$ is replaced in $\boldsymbol{x}$ by $\boldsymbol{x}_{\mathrm{d}}$, which contains equiprobable, zero mean, constant modulus, uncorrelated data symbols at the positions dictated by $\mathcal{I}_{\overline{L_h}}$. The expected value of the energy transmitted is given by:

$$\begin{aligned} E\left[\mathcal{E}^{\mathrm{sd}}_{\mathrm{QAM}}\right] &= \|\boldsymbol{x}_{L_h}\|^2 + \frac{1}{M}\boldsymbol{x}^H_{L_h}\boldsymbol{F}_{M\times\nu}\boldsymbol{F}^H_{M\times\nu}\boldsymbol{x}_{L_h} + \frac{1}{M}E\left(\boldsymbol{x}^H_{\mathrm{d}}\boldsymbol{F}_{M\times\nu}\boldsymbol{F}^H_{M\times\nu}\boldsymbol{x}_{\mathrm{d}}\right) \\ &= \mathcal{E}^{\mathrm{s}}_{\mathrm{QAM}} + \frac{1}{M}E\left(\boldsymbol{x}^H_{\mathrm{d}}\boldsymbol{F}_{M\times\nu}\boldsymbol{F}^H_{M\times\nu}\boldsymbol{x}_{\mathrm{d}}\right) \end{aligned} \tag{32}$$

The term $E\left(\boldsymbol{x}^H_{L_h}\boldsymbol{F}_{M\times\nu}\boldsymbol{F}^H_{M\times\nu}\boldsymbol{x}_{\mathrm{d}}\right)$ vanishes under a zero mean assumption on the data symbols. Notice that, again to be fair, we do not take the power of the data $E\left[\|\boldsymbol{x}_{\mathrm{d}}\|^2\right]$ into account. However, the term $E\left(\boldsymbol{x}^H_{\mathrm{d}}\boldsymbol{F}_{M\times\nu}\boldsymbol{F}^H_{M\times\nu}\boldsymbol{x}_{\mathrm{d}}\right)$ has to be included, since it represents the energy in the CP section due to the data, used by the receiver to eliminate the (intercarrier and intersymbol) interference at all, *data and pilot*, positions. In view of our assumption of uncorrelated data, we obtain:

$$\begin{aligned} E\left(\boldsymbol{x}^H_{\mathrm{d}}\boldsymbol{F}_{M\times\nu}\boldsymbol{F}^H_{M\times\nu}\boldsymbol{x}_{\mathrm{d}}\right) &= \sum_{i,j} E\left(\boldsymbol{x}_{\mathrm{d},i}\boldsymbol{x}^*_{\mathrm{d},j}\right)\left(\boldsymbol{F}_{M\times\nu}\boldsymbol{F}^H_{M\times\nu}\right)_{i,j} = \mathcal{E}^{\mathrm{QAM}}_x \sum_{i\in\mathcal{I}_{\overline{L_h}}}\left(\boldsymbol{F}_{M\times\nu}\boldsymbol{F}^H_{M\times\nu}\right)_{i,i} \\ &= \mathcal{E}^{\mathrm{QAM}}_x (M-L_h)(L_h-1) \end{aligned} \tag{33}$$

The power ratio in this case is defined as:

$$\mathrm{TPR}^{\mathrm{sd,s}}_{\mathrm{QAM}} = \frac{E\left[\mathcal{E}^{\mathrm{sd}}_{\mathrm{QAM}}\right]}{\mathcal{E}^{\mathrm{s}}_{\mathrm{QAM}}} = 1 + \frac{(M-L_h)(L_h-1)}{ML_h} > 1 \tag{34}$$

and hence:

$$\mathrm{MSE}^{\mathrm{s}}_{\mathrm{QAM}} = \frac{\mathrm{MSE}^{\mathrm{sd}}_{\mathrm{QAM}}}{\mathrm{TPR}^{\mathrm{sd,s}}_{\mathrm{QAM}}} < \mathrm{MSE}^{\mathrm{sd}}_{\mathrm{QAM}}$$

## VIII. OFDM/OQAM Sparse Preamble vs. CP-OFDM Sparse Preamble

From the previous analysis, it follows that the sparse preamble is generally the best choice for the preamble structure, in both OFDM systems. Let us then compare the estimation performances of the two systems, when using optimal sparse preambles, and with the same transmitted power for training. Clearly, in both cases the same model for the received signal, eq. (2), will hold. Moreover, if the spacing of the pilots, $M/L_h$, is large enough (theoretically equal to or larger than 2), the noise components at the corresponding outputs of the AFB for the OFDM/OQAM system will be uncorrelated. If we do not equalize the powers at the SFB outputs, the two MSE's will obviously be related as $\mathrm{MSE}^{\mathrm{s}}_{\mathrm{OQAM}} = \frac{\mathcal{E}^{\mathrm{QAM}}_x}{\mathcal{E}^{\mathrm{OQAM}}_x}\ \mathrm{MSE}^{\mathrm{s}}_{\mathrm{QAM}}$. Defining the power ratio for the two systems, $\text{TPR}^{\text{s}}_{\text{QAM/OQAM}}$, as the ratio of the training power in CP-OFDM over that in OFDM/OQAM, and scaling the output of the OFDM/OQAM SFB by $\sqrt{\text{TPR}^{\text{s}}_{\text{QAM/OQAM}}}$, we end up with:

$$\text{MSE}^{\text{s}}_{\text{OQAM}} = \frac{\mathcal{E}_x^{\text{QAM}}}{\mathcal{E}_x^{\text{OQAM}}} \frac{\text{MSE}^{\text{s}}_{\text{QAM}}}{\text{TPR}^{\text{s}}_{\text{QAM/OQAM}}}$$

We have previously seen that

$$\mathcal{E}^{\text{s}}_{\text{OQAM}} = L_h \mathcal{E}_x^{\text{OQAM}}$$

and

$$\mathcal{E}^{\text{s}}_{\text{QAM}} = L_h \mathcal{E}_x^{\text{QAM}}$$

The OFDM/OQAM sparse preamble generates $L_g$ nonzero samples at the output of the SFB, while the CP-OFDM sparse preamble yields $M + \nu = M + L_h - 1$ samples. The sampling rate at the output of the SFB's is the same for both systems. Hence, to equalize the energies per time unit for the two schemes, we have to form the power ratio as follows:

$$\text{TPR}^{\text{s}}_{\text{QAM/OQAM}} = \frac{\frac{1}{M+L_h-1} L_h \mathcal{E}_x^{\text{QAM}}}{\frac{1}{L_g} L_h \mathcal{E}_x^{\text{OQAM}}} = \frac{\frac{\mathcal{E}_x^{\text{QAM}}}{\mathcal{E}_x^{\text{OQAM}}} L_g}{M + L_h - 1} \tag{35}$$

and finally

$$\text{MSE}^{\text{s}}_{\text{OQAM}} = \frac{M + L_h - 1}{L_g} \text{MSE}^{\text{s}}_{\text{QAM}} \tag{36}$$

For example, let $L_h = 32$. Then, for $L_g = M$, the CP-OFDM sparse preamble turns out to be superior to the corresponding OFDM/OQAM sparse preamble, while for $L_g = KM$, with $2 \leq K \leq 5$, the OFDM/OQAM sparse preamble is approximately $3 - 9$ dB better.

*Remarks.*

1) Note that $(M + L_h - 1)/L_g$ is the *ratio of the time durations* of the transmit pulses employed by the two systems.
2) The performance difference can be even greater if we want to achieve a lower PAPR in the CP-OFDM system. We will then have to use *unequal* equipowered pilots, which leads to a slightly worse performance of the CP-OFDM sparse preamble.
3) Nevertheless, *at the cost of increasing the bandwidth in the OFDM/OQAM system*, the OFDM/OQAM and CP-OFDM sparse preambles can become MSE equivalent, in the following way. Note that, due to the good time localization of the OFDM/OQAM pulse, there is always a subinterval of the total pulse duration in the OFDM/OQAM system with the same length as the CP-OFDM modulator output, that carries almost all of the energy of the pulse. In view of the even symmetry of $g$, we can consider the subinterval $[-\lceil (M + L_h - 1)/2 \rceil, \lceil (M + L_h - 1)/2 \rceil]$ around its center, where $\lceil a \rceil$ denotes the smallest integer that is not smaller than $a$. Then, for practical values of $M, L_h$, it can be easily verified than:

$$\sum_{l=-\lceil (M+L_h-1)/2 \rceil}^{\lceil (M+L_h-1)/2 \rceil} g^2(\lceil L_g/2 \rceil + l) \approx 0.99$$

Therefore, we only need to observe this interval to approximately reconstruct the preamble vector at the receiver. Then the transmit pulses in the two systems have approximately the same duration (albeit with OQAM bandwidth increased), thus leading to almost the same MSE performance for the two sparse preambles. Again, the OFDM/QAM sparse preamble can be slightly better if we use unequal equipowered pilots in the CP-OFDM sparse preamble for a lower PAPR.

4) This last comparison setup does not affect any of the previous results, since we have always compared preamble structures for the same system and therefore the same pulse duration. We only need to be careful in the last two scenarios of Section VII for the OFDM/OQAM system. For Scenario 2, it can be seen that, for sufficiently large $L_g$,

$$\mathrm{TPR}_{\mathrm{OQAM}}^{\mathrm{sd,s}} \approx \frac{(M+L_h-1)(1+\zeta)}{M+L_h-1+\frac{M}{2}}$$

The same result also holds in Scenario 3. This shows that the last comparison setup reduces the MSE differences in the sparse-data case.

## IX. Error Floor Analysis for OFDM/OQAM and CP-OFDM systems

The fact that the intrinsic interference is a part of the error signal dependent on transmit signal components, indicates the existence of an error floor behavior of the perfromance curves for the OFDM/OQAM system. In CP-OFDM , there is not such a problem due to the orthogonality of the DFT transformation and the use of the CP. This orthogonality eliminates the interferences coming from the neighboring symbols on each pilot symbol. However, the OFDM/OQAM system possesses orthogonality only in the real field. Therefore, interferences to each subcarrier symbol coming from the neighboring subcarrier symbols are inevitable in the presence of a complex CFR. Generally speaking, the interference is minimized for large $M$ and small $K$. Large $M$ leads to better localization in the frequency domain, while small $K$ minimizes the number of overlapping OQAM vector symbols in the temporal direction. We will analyze the error floor behavior of both systems to prove the aforementioned claims.

First, note that the process of estimating the channel in $N \geq L_h$ ($N \leq M$) positions in the frequency domain, then finding the CIR by translating these estimates to the time domain, and finally obtaining the channel gains at all subcarriers through a DFT operation, is essentially a DFT interpolation of the original frequency domain estimates. The CFR coefficients, originally estimated through the LS estimator, can be expressed as $\boldsymbol{H}_N = \boldsymbol{F}_{N\times L_h}\boldsymbol{h}$, where $\boldsymbol{F}_{N\times L_h}$ is the $N \times L_h$ submatrix of the DFT matrix consisting of its first $L_h$ columns and its $N$ rows corresponding to the indices of the frequency domain channel gains we wish to estimate. The final estimates of the frequency domain channel gains are therefore given by $\hat{\boldsymbol{H}} = \boldsymbol{F}_{M\times L_h}\left(\boldsymbol{F}_{N\times L_h}^H \boldsymbol{F}_{N\times L_h}\right)^{-1}\boldsymbol{F}_{N\times L_h}^H \hat{\boldsymbol{H}}_N$. Assuming that $N$ is a divisor of $M$ and the $N$ pilots are equispaced, the last expression becomes $\hat{\boldsymbol{H}} = \frac{1}{N}\boldsymbol{F}_{M\times L_h}\boldsymbol{F}_{N\times L_h}^H \hat{\boldsymbol{H}}_N$.

Consider now the received signal on the $m$th subcarrier for the OFDM/OQAM preamble:

$$y_m = H_m a_m + \jmath \boldsymbol{\mathcal{H}}_m^T \boldsymbol{u}_m + \eta_m$$

We have dropped the temporal index for notation simplification. Here $a_m$ is a real symbol (equal for example to $\pm\sqrt{\mathcal{E}_x^{\mathrm{OQAM}}}$ for a QPSK constellation), $\boldsymbol{u}_m$ is an $(N-1)\times 1$ vector of crosscorrelations of pulses transmitted on the active subcarriers with the pulse transmitted on the $m$th subcarrier, incorporating the corresponding symbols as well, and $\boldsymbol{\mathcal{H}}_m$ an $(N-1)\times 1$ vector of the channel gains on these subcarriers. There are two ways to estimate the original channel gain: Either as $\hat{H}_m = \frac{y_m}{a_m} = H_m + \jmath\frac{\boldsymbol{\mathcal{H}}_m^T\boldsymbol{u}_m}{a_m} + \frac{\eta_m}{a_m}$ or as $\hat{H}_m = \frac{y_m}{a_m+\jmath s_m} = H_m\frac{a_m}{a_m+\jmath s_m} + \jmath\frac{\boldsymbol{\mathcal{H}}_m^T\boldsymbol{u}_m}{a_m+\jmath s_m} + \frac{\eta_m}{a_m+\jmath s_m}$, where $s_m$ is the sum of the entries of $\boldsymbol{u}_m$ coming from the immediately adjacent subcarriers of $m$. The second way to estimate the channel comes from a usual assumption in the OFDM/OQAM system, namely that the channel gain can be considered to be constant in the first-order neighborhood of $m$, especially if $M$ is large and $K$ is small. In this case, the received model can be approximated by $y_m \approx H_m(a_m + \jmath s_m) + \eta_m$, which justifies the second way of estimating the desired channel gain. In our analysis, and for the case that $M/N \geq 2$, we will use the first way of estimating the channel, because we do not have to assume anything about first-order neighborhoods or other approximations. In this way, our analysis becomes exact.

Assuming that the noise is zero mean and its variance is $\sigma^2$, we obtain $\mu_m = E[\hat{H}_m] = H_m + \jmath\frac{\boldsymbol{\mathcal{H}}_m^T\boldsymbol{u}_m}{a_m}$, where the channel is considered to be an unknown but otherwise deterministic quantity and the training symbols deterministic quantities. Here $E[\cdot]$ denotes the expectation operator w.r.t. the noise statistics. The MSE for the above estimate is given by:

$$\text{MSE}_m = E[|\hat{H}_m - H_m|^2] = \frac{|\boldsymbol{\mathcal{H}}_m^T\boldsymbol{u}_m|^2}{a_m^2} + \frac{\sigma^2}{a_m^2} \tag{37}$$

The last equation justifies the existence of an error floor for the OFDM/OQAM system, since, as $\sigma^2 \longrightarrow 0$, $\text{MSE}_m \longrightarrow \frac{|\boldsymbol{\mathcal{H}}_m^T\boldsymbol{u}_m|^2}{a_m^2}$, i.e., as the SNR increases, the intrinsic interference becomes a dominant phenomenon.

On the contrary, for the CP-OFDM system, the received signal model is $y_m = H_m a_m + \eta_m$ and $\hat{H}_m = \frac{y_m}{a_m} = H_m + \frac{\eta_m}{a_m}$. Thus the estimate is obviously unbiased and its MSE, given by $\sigma^2/a_m^2$, tends to zero as the SNR increases.

For the case that $M/N = 1$, i.e., for a full preamble, we will use the second estimate for the OFDM/OQAM system. This estimate leads to better performance as it has been shown in [13]. The increase of the magnitude of the pseudo-pilot, $a_m + \jmath s_m$, as opposed to $a_m$ compensates for the inaccuracy introduced by considering the channel to be constant in every first-order neighborhood, especially in the SNR regime where realistic systems operate. With this estimating method, the mean value of the exact estimate is $\mu_m = E[\hat{H}_m] = H_m\frac{a_m}{a_m+\jmath s_m} + \jmath\frac{\boldsymbol{\mathcal{H}}_m^T\boldsymbol{u}_m}{a_m+\jmath s_m}$ and its MSE $\text{MSE}_m = |H_m|^2\left|\frac{a_m}{a_m+\jmath s_m} - 1\right|^2 + \frac{|\boldsymbol{\mathcal{H}}_m^T\boldsymbol{u}_m|^2}{a_m^2+s_m^2} + \frac{\sigma^2}{a_m^2+s_m^2} + 2\frac{a_m}{a_m^2+s_m^2}\Re\left\{-\jmath H_m^*\boldsymbol{\mathcal{H}}_m^T\boldsymbol{u}_m\right\}$.

We may now stack the $N$ estimates of the channel gains for the OQAM system to obtain the $N\times 1$ vector $\hat{\boldsymbol{H}}_N = \boldsymbol{H}_N + \boldsymbol{w}_1 + \boldsymbol{w}_2$.

<u>Case 1:</u> $M/N \geq 2$

We then have $\boldsymbol{w}_1 = \left[\jmath\frac{\boldsymbol{\mathcal{H}}_{i_0}^T\boldsymbol{u}_{i_0}}{a_{i_0}}, \jmath\frac{\boldsymbol{\mathcal{H}}_{i_1}^T\boldsymbol{u}_{i_1}}{a_{i_1}}, \ldots, \jmath\frac{\boldsymbol{\mathcal{H}}_{i_{N-1}}^T\boldsymbol{u}_{i_{N-1}}}{a_{i_{N-1}}}\right]^T$ and $\boldsymbol{w}_2 = \left[\frac{\eta_{i_0}}{a_{i_0}}, \frac{\eta_{i_1}}{a_{i_1}}, \ldots, \frac{\eta_{i_{N-1}}}{a_{i_{N-1}}}\right]$, where $i_0 \in \{0, 1, \ldots, M/N-1\}$. It can be easily shown that:

$$\text{MSE}_N \;=\; E\left[\|\hat{\boldsymbol{H}} - \boldsymbol{H}\|^2\right] = \frac{1}{N^2}\|\boldsymbol{F}_{M\times L_h}\boldsymbol{F}_{N\times L_h}^H\boldsymbol{w}_1\|^2 + \frac{1}{N^2}\sigma^2\|\boldsymbol{F}_{M\times L_h}\boldsymbol{F}_{N\times L_h}^H\boldsymbol{w}_3\|^2$$

where $\boldsymbol{w}_3 = \left[\frac{1}{a_{i_0}}, \frac{1}{a_{i_1}}, \ldots, \frac{1}{a_{i_{N-1}}}\right]$.

With $\sigma^2 \longrightarrow 0$, the error floor results:

$$\mathrm{MSE}_N^{\mathrm{floor}} = \frac{1}{N^2}\|\boldsymbol{F}_{M\times L_h}\boldsymbol{F}_{N\times L_h}^H\boldsymbol{w}_1\|^2$$

<u>Case 2:</u> $M/N = 1$

This is the case of the full preamble. Then:

$$\boldsymbol{w}_1 = \Bigg[-\jmath H_0\frac{s_0}{a_0+\jmath s_0}+\jmath\frac{\boldsymbol{\mathcal{H}}_0^T\boldsymbol{u}_0}{a_0+\jmath s_0}, -\jmath H_1\frac{s_1}{a_1+\jmath s_1}+\jmath\frac{\boldsymbol{\mathcal{H}}_1^T\boldsymbol{u}_1}{a_1+\jmath s_1}, \ldots, -\jmath H_{M-1}\frac{s_{M-1}}{a_{M-1}+\jmath s_{M-1}}+\jmath\frac{\boldsymbol{\mathcal{H}}_{M-1}^T\boldsymbol{u}_{M-1}}{a_{M-1}+\jmath s_{M-1}}\Bigg]^T$$

and $\boldsymbol{w}_2 = \left[\frac{n_0}{a_0+\jmath s_0}, \frac{n_1}{a_1+\jmath s_1}, \ldots, \frac{n_{M-1}}{a_{M-1}+\jmath s_{M-1}}\right]$. Now:

$$\mathrm{MSE}_M = \frac{1}{M^2}\|\boldsymbol{F}_{M\times L_h}\boldsymbol{F}_{M\times L_h}^H\boldsymbol{w}_1\|^2 + \frac{1}{M^2}\sigma^2\|\boldsymbol{F}_{M\times L_h}\boldsymbol{F}_{M\times L_h}^H\boldsymbol{w}_3\|^2$$

where $\boldsymbol{w}_3 = \left[\frac{1}{a_0+\jmath s_0}, \frac{1}{a_1+\jmath s_1}, \ldots, \frac{1}{a_{M-1}+\jmath s_{M-1}}\right]$. Thus, the error floor is given in this case by:

$$\mathrm{MSE}_M^{\mathrm{floor}} = \frac{1}{M^2}\|\boldsymbol{F}_{M\times L_h}\boldsymbol{F}_{\times L_h}^H\boldsymbol{w}_1\|^2$$

*Remarks:*

1) There is one more source of error floor generation. This is the channel length. If the channel length is too large, then the received signal models for the OFDM/OQAM system used in this paper and in the literature do not hold any more. We do not assume such a degenerate case in our analysis or in the simulation section.

2) The above analysis holds only for sample-spaced channels. There is an extra floor generating mechanism if the channel is nonsample-spaced.

## X. Simulations

In this section, we present simulation results to verify our analysis. The channel follows the veh-A model [2]. The CIR is initially generated with 29 taps and then zero padded to the closest power of two, that is, $L_h = 32$ taps. We plot the normalized MSE (NMSE), i.e., $E(\|\boldsymbol{H} - \hat{\boldsymbol{H}}\|^2/\|\boldsymbol{H}\|^2)$, versus the transmit bit SNR ($E_b/N_0$). The curves are the result of averaging 200 channel realizations. For each channel realization, 300 different noise realizations are considered. QPSK modulation is employed.

### A. CP-OFDM

The results are for $M = 1024$ subcarriers and a CP length $\nu = 31$. Fig. 1a shows the NMSE performance of the CP-OFDM system for the full and sparse preambles, where for the full preamble we use the CFR estimates as in (3). We observe that the performance of a sparse preamble with $L_h$ equispaced and equal pilot tones is much better. Note that the difference of the performances in theory should be $10\log_{10}(M/L_h) = 15.05$ dB, which can be seen to agree with the simulation. This difference is independent of the SNR value, which justifies the fact that the curves are parallel. Fig. 1b depicts the result of the processing described in Section V-C. As expected, the two preambles lead then to the same performance.

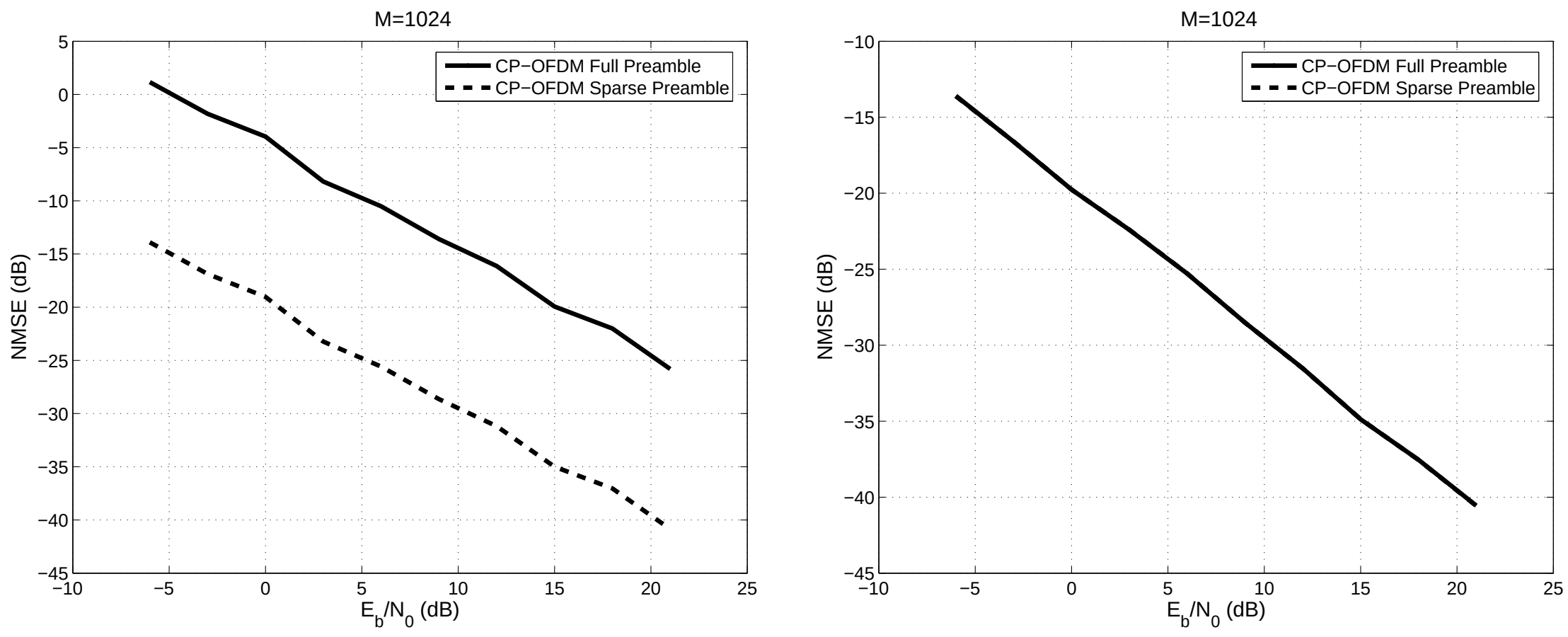

(a) (b)

Fig. 1. NMSE performance of the CP-OFDM system for the full vs. sparse preamble case: (a) directly measuring the performance in the frequency domain; (b) after applying the frequency→time→frequency processing.

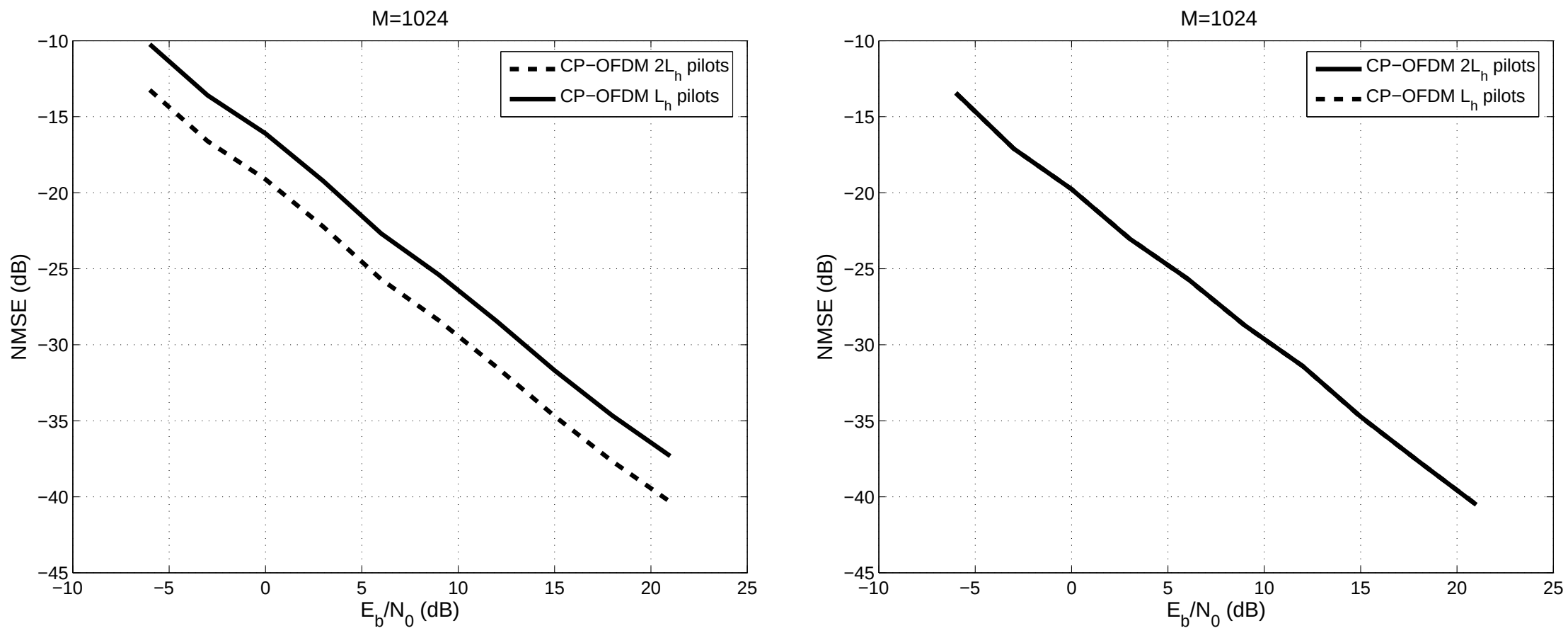

(a) (b)

Fig. 2. NMSE performance of the CP-OFDM system for the sparse preamble with $2L_h$ pilots vs. the sparse preamble with $L_h$ pilots: (a) before the power equalization; (b) after the power equalization.

Fig. 2 presents the performance of the sparse preamble with $L_h$ pilots versus a sparse preamble with $P = 2L_h$ pilots, before and after the power equalization. The latter preamble performs better before the power equalization, since it leads to the transmission of more power. After the equalization of the powers, the two preambles perform similarly, as expected.

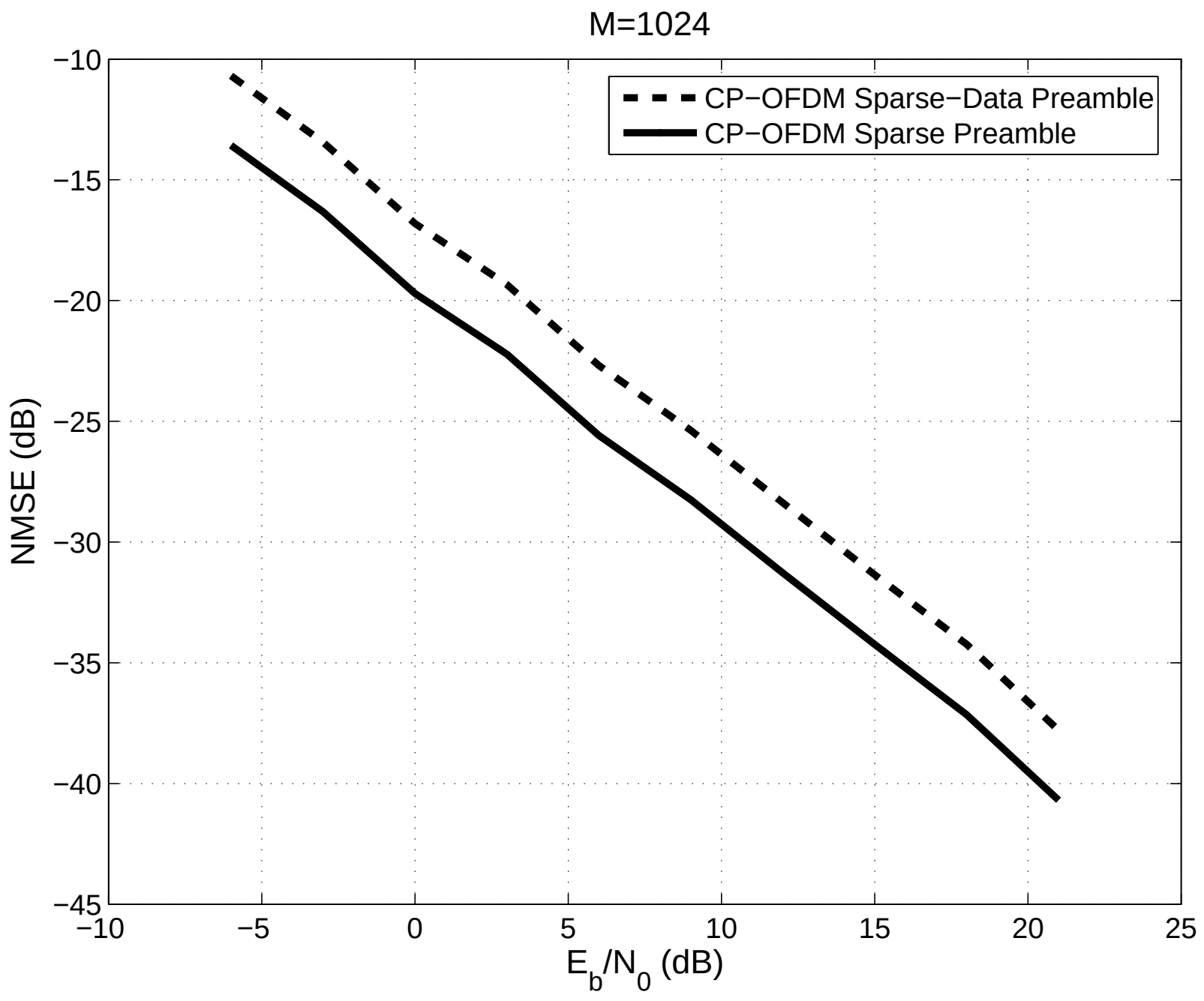

Fig. 3. NMSE performance of the CP-OFDM system for the sparse-data vs. the sparse preamble case.

In Fig. 3, the sparse and sparse-data preambles are compared. We observe that the sparse preamble is better when the transmit power is the same in both cases. The difference between the curves can be easily checked to approximately follow the theoretical results. For example, for $M = 1024$ and $L_h = 32$, the theoretical value of the performance difference is 2.87 dB, which shows up in the figure.

### B. OFDM/OQAM

For the OFDM/OQAM system, and using filter banks given in [17], [4], Fig. 1 translates to Fig. 4. The differences between the curves before and after the power equalization can be easily checked to be in accordance with the analytical results. For $M = 1024, L_h = 32, K = 4$ and the adopted pulse $g$, the theoretical difference before the power equalization is $10 \log_{10} \{M/[L_h(1 + 2\beta)]\} \approx 12.5$ dB, which can be seen in Fig. 4a. In Fig. 4b, the performances are similar, verifying the result proved in Appendix V. The performance of the sparse preamble with $P = 2L_h$ pilots is compared to that of the sparse preamble with $L_h$ pilots in Fig. 5. For the mixed sparse-data case, we choose to implement Scenario 3 as described in Section VII-A.2. This is the most involved among the sparse-data scenarios in the OFDM/OQAM system and an example for this is provided in Fig. 6. Note the error floor in the sparse-data scenario.

### C. Comparison

The sparse preambles for the CP-OFDM and OFDM/OQAM systems are compared in Fig. 7. The superior-ity of the OFDM/OQAM sparse preamble, when the entire transmit pulse duration is considered, is evident. The analytical results can be seen to be approximately verified. Thus, for Fig. 7a, the theoretical difference is $10\log_{10}\left[KM/(M+L_h-1)\right] \approx 4.5$ dB, while for Fig. 7b, it is approximately 5.9 dB. These values agree with the difference of the experimental curves. The two systems, however, perform similarly in the alternative comparison setup described in Section VIII (Remark 3), as shown in Fig. 8.

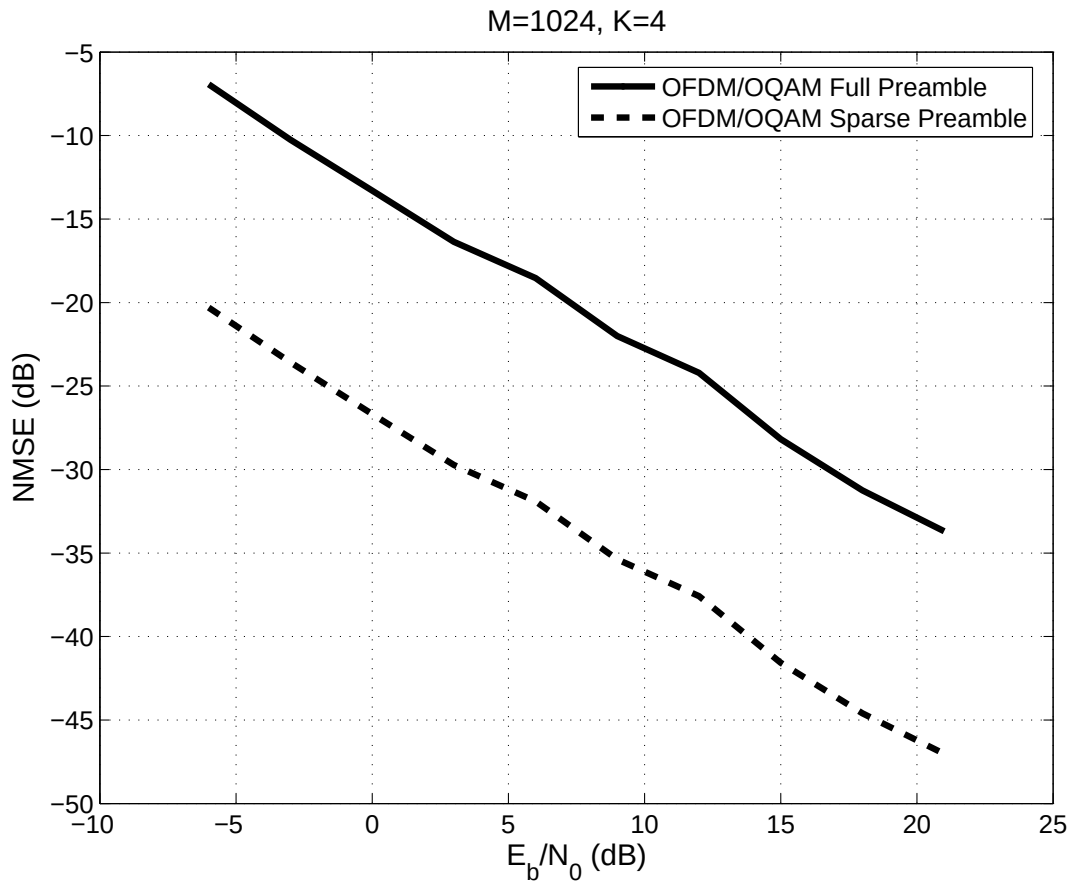

(a)

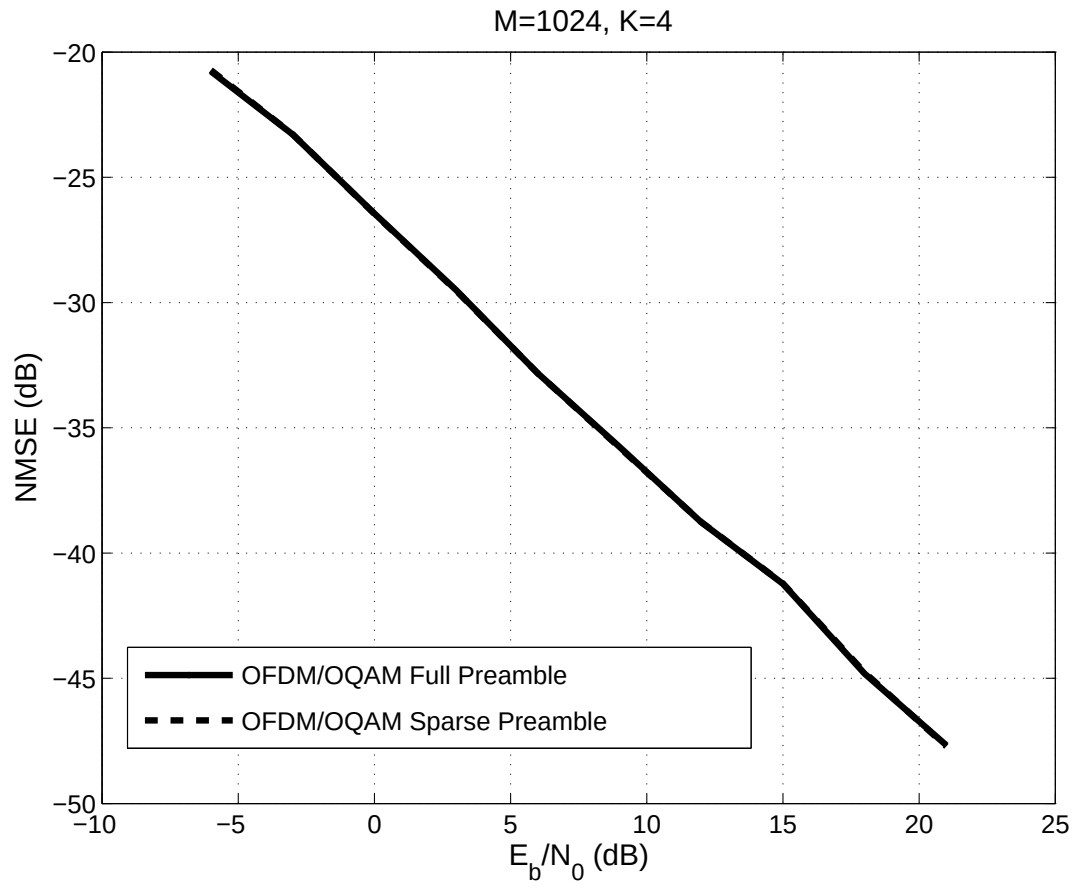

(b)

Fig. 4. NMSE performance of the OFDM/OQAM system for the full vs. sparse preamble case: (a) directly measuring the performance in the frequency domain; (b) after applying the frequency→time→frequency processing.

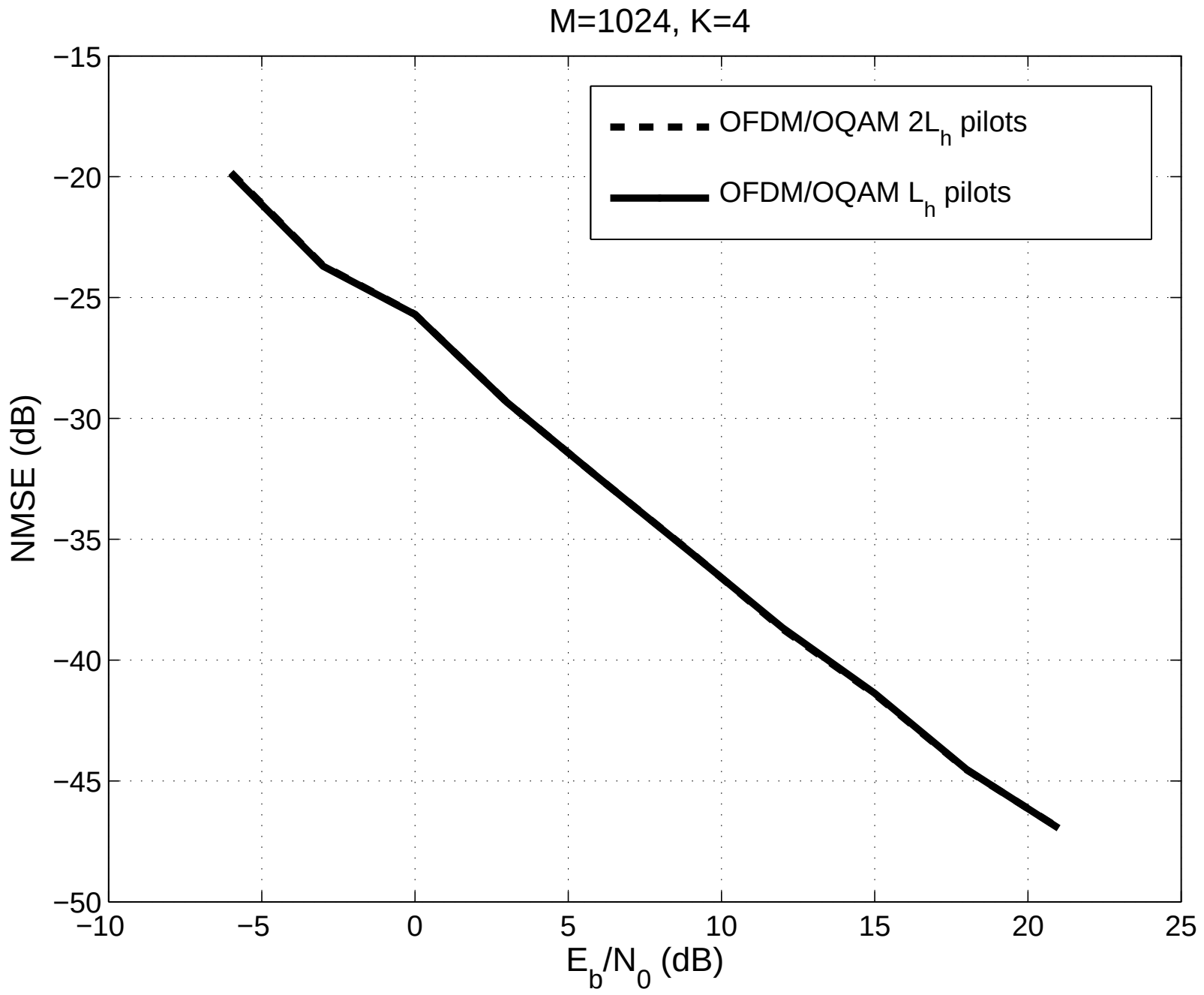

Fig. 5. NMSE performance of the OFDM/OQAM system for the sparse preamble with $P = 2L_h$ pilots vs. $L_h$ pilots, with power equalization.

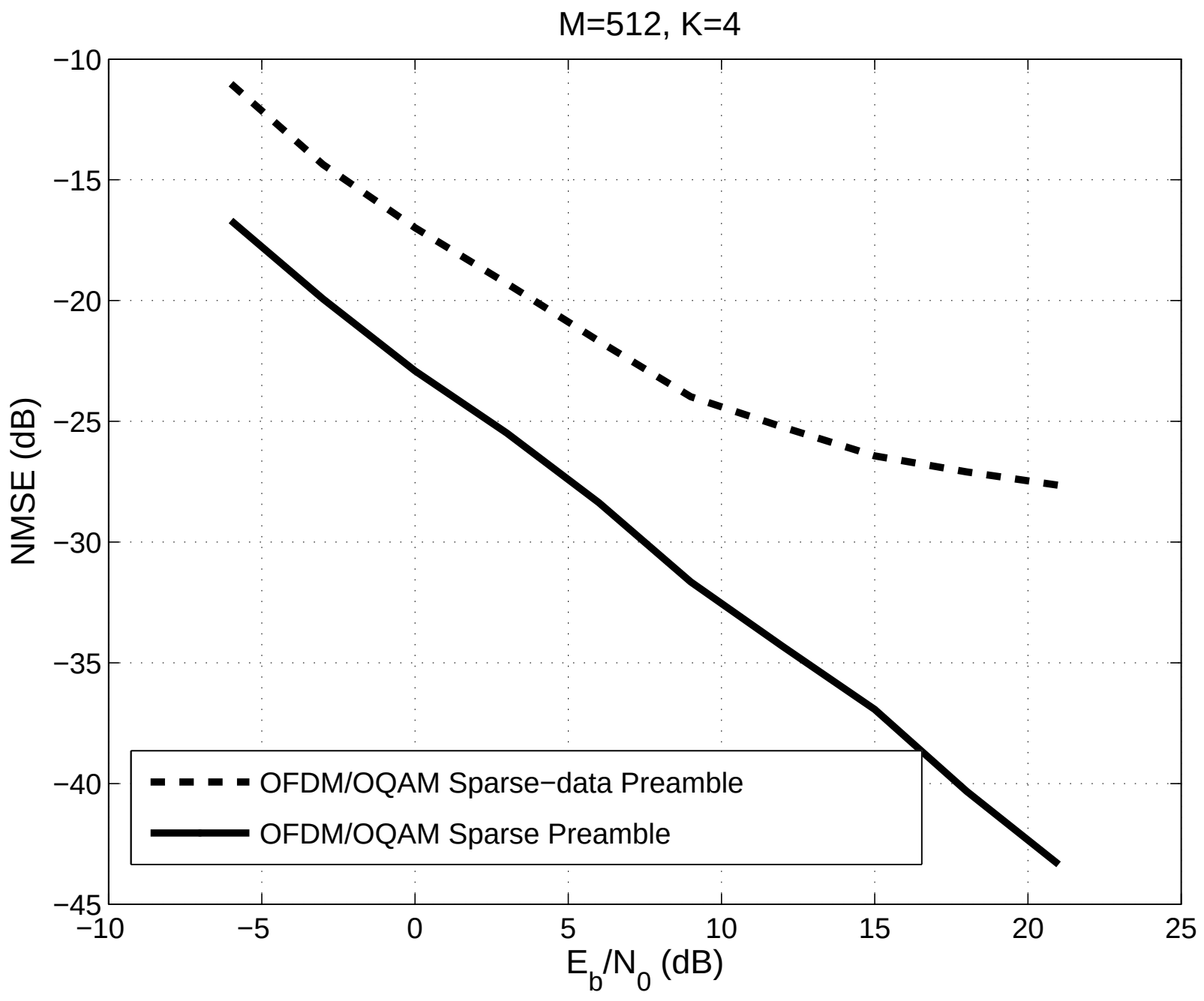

Fig. 6. NMSE performance of the OFDM/OQAM system for the sparse-data vs the sparse preamble, with power equalization.

## XI. Conclusions

Optimal preamble design for LS channel estimation in CP-OFDM and OFDM/OQAM systems was addressed in this paper, for both full and sparse preambles. In contrast to earlier related work on CP-OFDM, the energy spent for the CP transmission was also taken into account when assessing the energy budget for training. This turned out to lead to the requirement of *equal* instead of simply equipowered pilot tones for the CP-OFDM sparse preamble. Equipowered and equispaced pilot tones were shown to comprise the optimal sparse preamble for OFDM/OQAM. Possible gains from loading data on the inactive subcarriers of a sparse preamble were also investigated. The sparse preamble with as many pilot tones as channel taps turned out to be generally the best choice in terms of both estimation performance and economy. The OFDM/OQAM optimal sparse preamble was compared with that of CP-OFDM and shown to allow for a significantly better performance, provided the whole pulse is transmitted when training. Nevertheless, it will perform similarly to CP-OFDM, at the cost of bandwidth expansion, if the tails of the (well time-localized) pulse are left out in the transmission of the preamble. Apart from the bandwidth difference between the two systems, this shows the fundamental *structural* similarity of the two systems. Our analytical results were confirmed via simulations.

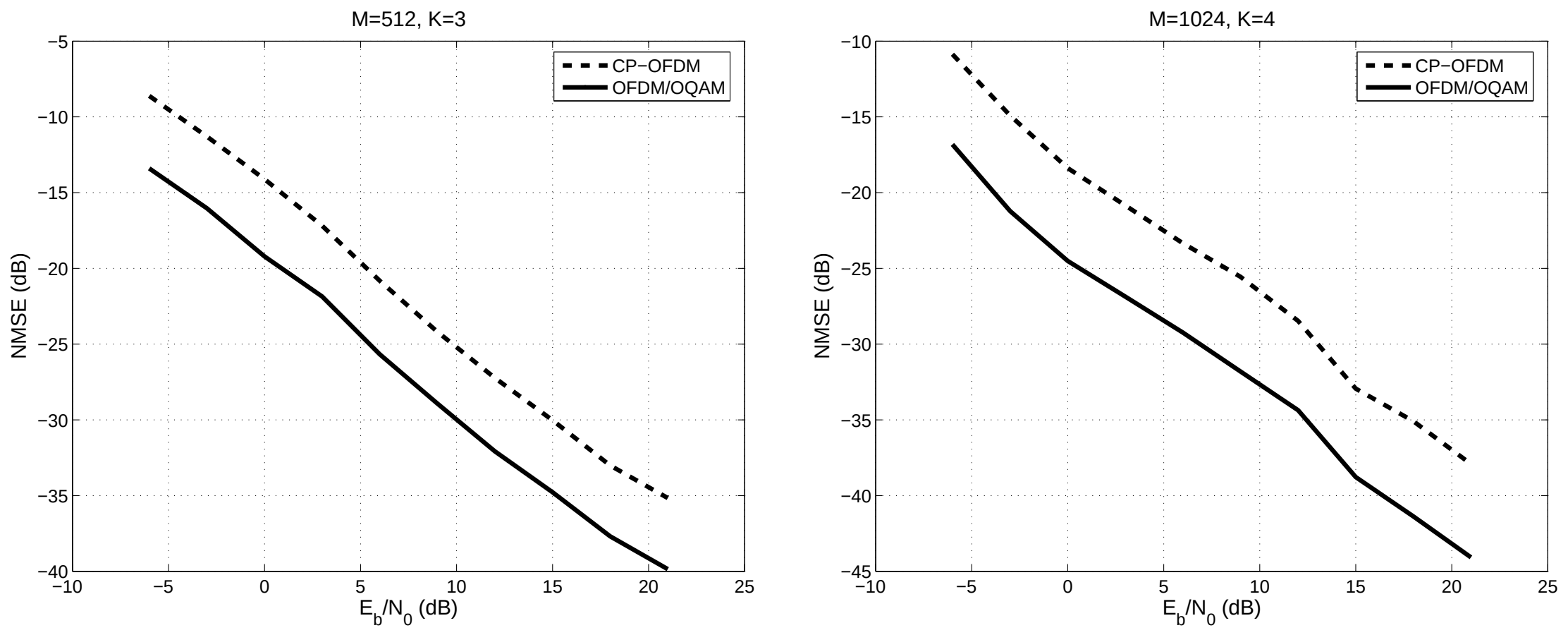

(a) (b)

Fig. 7. NMSE performance of the CP-OFDM and OFDM/OQAM sparse preambles: (a) $M = 512$, $K = 3$; (b) $M = 1024$, $K = 4$.

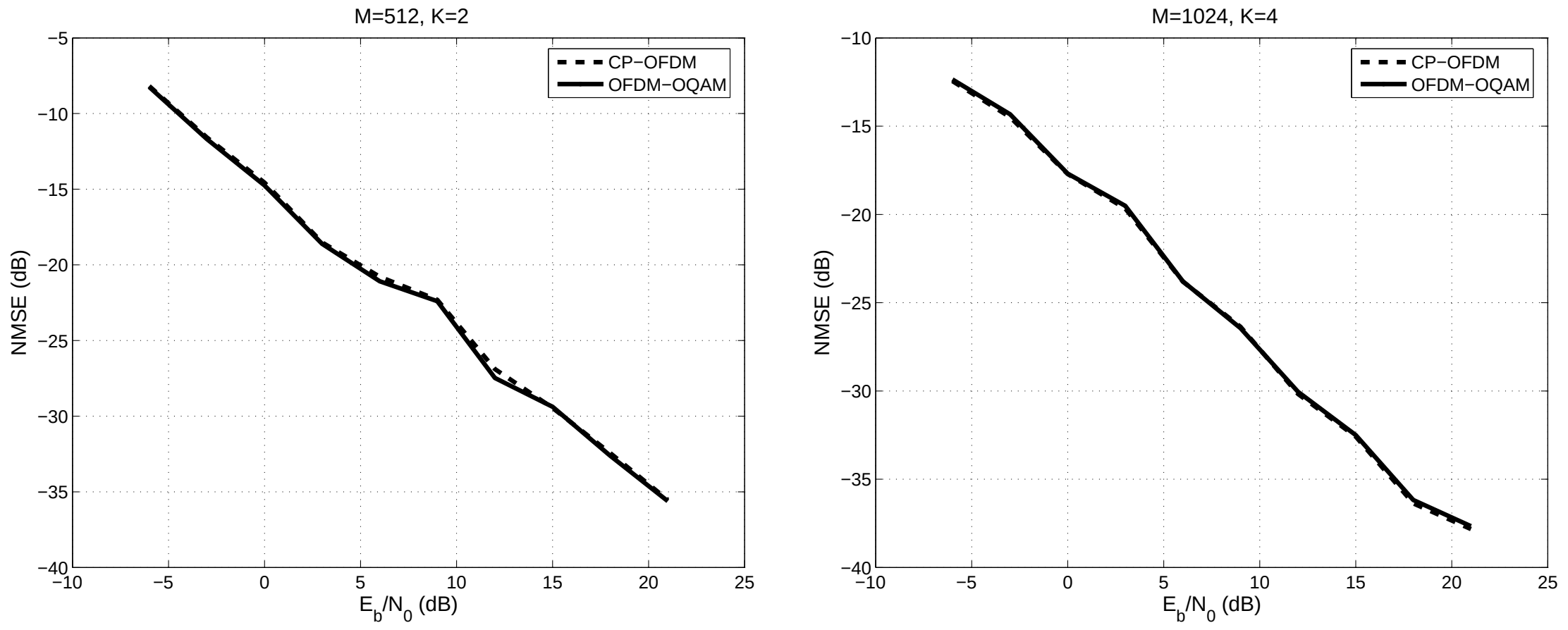

(a) (b)

Fig. 8. NMSE performance of the CP-OFDM and OFDM/OQAM sparse preambles with the alternative comparison setup: (a) $M = 512$, $K = 2$; (b) $M = 1024$, $K = 4$.

## Appendix I

## CP-OFDM: Sparse Preamble With Equispaced and Equal Pilot Tones

It was proved in [19], [3] that a sparse preamble of $L_h$ pilots is MSE-optimal subject to a training energy constraint when it is built with equipowered and equispaced pilot tones. That energy constraint did not include the energy spent for CP. Our goal here is to determine the optimal sparse preamble when the training energy includes the CP energy as well.

Let $\mathcal{I}_{L_h} = \{i_0, i_1, \dots, i_{L_h-1}\}$ be the set of indices of the nonzero pilot tones in the sparse preamble, and denote by $\mathcal{I}_{\overline{L_h}}$ the set $\{0, 1, \dots, M-1\} \setminus \mathcal{I}_{L_h}$, consisting of the indices of the null tones. Stacking the $L_h$ CFR estimates in the vector $\hat{\boldsymbol{H}}_{L_h}$, we can find the CIR as $\hat{\boldsymbol{h}} = \boldsymbol{F}^{-1}_{L_h \times L_h} \hat{\boldsymbol{H}}_{L_h} = \boldsymbol{h} + \boldsymbol{F}^{-1}_{L_h \times L_h} \boldsymbol{\varepsilon}$, where $\boldsymbol{\varepsilon}$ is the $L_h \times 1$ vector with entries $\eta_m / x_m$, $m \in \mathcal{I}_{L_h}$, and $\boldsymbol{F}_{L_h \times L_h}$ is the $L_h \times L_h$ submatrix of the $M \times M$ DFT matrix $\boldsymbol{\mathcal{F}}$ consisting of its $L_h$ first columns and its rows corresponding to the indices in $\mathcal{I}_{L_h}$. Thus, the MSE of the above estimate is given by

$$\mathrm{MSE}_{L_h} = \mathrm{tr}\left[ \boldsymbol{\mathcal{C}}_{L_h} \left( \boldsymbol{F}_{L_h \times L_h} \boldsymbol{F}^H_{L_h \times L_h} \right)^{-1} \right], \tag{38}$$

where $\boldsymbol{\mathcal{C}}_{L_h} = \sigma^2 \mathrm{diag}\left( 1/|x_{i_0}|^2, 1/|x_{i_1}|^2, \dots, 1/|x_{i_{L_h-1}}|^2 \right)$ is the covariance of $\boldsymbol{\varepsilon}$. Without loss of generality, we can order the diagonal elements of $\boldsymbol{\mathcal{C}}_{L_h}$ in descending order, as follows $\frac{\sigma^2}{|x_{i_0}|^2} \geq \frac{\sigma^2}{|x_{i_1}|^2} \geq \cdots \geq \frac{\sigma^2}{|x_{i_{L_h-1}}|^2}$, and the eigenvalues of $\left( \boldsymbol{F}_{L_h \times L_h} \boldsymbol{F}^H_{L_h \times L_h} \right)^{-1}$ in ascending order, i.e., $\lambda_0 \leq \lambda_1 \leq \cdots \leq \lambda_{L_h-1}$. Then [**?**, Lemma 1]:

$$\mathrm{MSE}_{L_h} = \mathrm{tr}\left[ \boldsymbol{\mathcal{C}}_{L_h} \left( \boldsymbol{F}_{L_h \times L_h} \boldsymbol{F}^H_{L_h \times L_h} \right)^{-1} \right] \geq \sum_{m=0}^{L_h-1} \frac{\sigma^2}{|x_{i_m}|^2} \lambda_m \tag{39}$$

with equality if and only if the matrix in the brackets is diagonal. Let the eigenvalue decomposition (EVD) of $\left( \boldsymbol{F}_{L_h \times L_h} \boldsymbol{F}^H_{L_h \times L_h} \right)^{-1}$ be $\boldsymbol{U} \boldsymbol{\Lambda} \boldsymbol{U}^H$, with $\boldsymbol{U}$ being its $L_h \times L_h$ eigenvector matrix and $\boldsymbol{\Lambda} = \mathrm{diag}(\lambda_0, \lambda_1, \dots, \lambda_{L_h-1})$. To optimize the selection of $\boldsymbol{F}_{L_h \times L_h}$, we must find the optimal placement of the pilots in the training vector. This choice will determine $\boldsymbol{U}$ and $\boldsymbol{\Lambda}$. In order to satisfy (39) with equality, we have to make a placement that yields $\boldsymbol{U} = \boldsymbol{I}_{L_h}$. It is known [19], that a placement of the pilot tones with this property is the equidistant one. We will focus on this placement now to find its optimal pilot tones. In the next appendix, we show that the sparse preamble just obtained is the globally optimal one.

After the CP insertion, the energy reaching the transmit antenna is given by:

$$\begin{aligned} \|\boldsymbol{s}_{\mathrm{QAM}}\|^2 &= \|\boldsymbol{x}\|^2 + \frac{1}{M} \boldsymbol{x}^H \boldsymbol{F}_{M \times \nu} \boldsymbol{F}^H_{M \times \nu} \boldsymbol{x} \\ &= \sum_{m=0}^{L_h-1} |x_{i_m}|^2 + \frac{1}{M} \sum_{k,m=0}^{L_h-1} x^*_{i_k} x_{i_m} \sum_{l=1}^{\nu} e^{\jmath \frac{2\pi}{L_h}(m-k)(M-1-\nu+l)} \end{aligned}$$

Thus, the optimization problem we have to solve is stated as:

$$\min_{x_{i_m}, m \in \mathcal{I}^{\mathrm{e}}_{L_h}} \quad \frac{\sigma^2}{L_h} \sum_{m=0}^{L_h-1} \frac{1}{|x_{i_m}|^2} \tag{40}$$

$$\text{such that (s.t)} \sum_{m=0}^{L_h-1} |x_{i_m}|^2 + \frac{1}{M} \sum_{k,m=0}^{L_h-1} x^*_{i_k} x_{i_m} \sum_{l=1}^{\nu} e^{\jmath \frac{2\pi}{L_h}(m-k)(M-1-\nu+l)} \leq \mathcal{E} \tag{41}$$

where $\mathcal{I}^{\mathrm{e}}_{L_h}$ is any (yet fixed) of the $M/L_h$ equispaced placements of the pilot tones and $\mathcal{E}$ the total transmit energy available for training.

*Proposition 1:* An optimal solution for the problem (40), (41) is given by equal symbols. This yields a local minimum of the constrained problem.

*Proof:* Forming the Lagrangian function for the above problem, we get:

$$\mathcal{J}(x_{i_0}, x_{i_1}, \ldots, x_{i_{L_h-1}}) = \frac{\sigma^2}{L_h}\sum_{m=0}^{L_h-1}\frac{1}{|x_{i_m}|^2} + \mu\left[\sum_{m=0}^{L_h-1}|x_{i_m}|^2 + \frac{1}{M}\sum_{k,m=0}^{L_h-1}x_{i_k}^* x_{i_m}\sum_{l=1}^{\nu}e^{\jmath\frac{2\pi}{L_h}(m-k)(M-1-\nu+l)} - \mathcal{E}\right]$$

where $\mu$ is the Lagrange multiplier.[7] Setting the gradient of $\mathcal{J}$ to zero, we obtain:

$$-\frac{\sigma^2}{L_h}\frac{x_{i_m}^*}{|x_{i_m}|^4} + \mu\left[x_{i_m}^* + \frac{1}{M}\sum_{k=0}^{L_h-1}x_{i_k}^*\sum_{l=1}^{\nu}e^{\jmath\frac{2\pi}{L_h}(m-k)(M-1-\nu+l)}\right] = 0, \quad m = 0, 1, \ldots, L_h - 1$$

Multiplying by $x_{i_m}$ and summing over $m$, $\mu$ can be found as $\mu = \frac{\sigma^2}{\mathcal{E}L_h}\sum_{m=0}^{L_h-1}\frac{1}{|x_{i_m}|^2}$, where we have used the constraint with equality.

Consider the preamble vector with *equal* training symbols. Then by the energy constraint we can easily find that $|x_{i_m}|^2 = |x|^2 = \frac{\mathcal{E}}{L_h}$, since the CP energy part vanishes (cf. (24)). We can check that the Lagrange equations are jointly satisfied by this point:

$$\left.\frac{\partial\mathcal{J}}{\partial x_{i_m}}\right|_{(x,x,\ldots,x)} = -\frac{\sigma^2}{L_h|x|^4}x^* + \frac{\sigma^2}{\mathcal{E}}\frac{1}{|x|^2}x^* = \frac{\sigma^2}{|x|^2}x^*\left(\frac{1}{\mathcal{E}} - \frac{1}{L_h|x|^2}\right) = 0$$

Furthermore, it can be easily checked that the Hessian in a neighborhood of $(x, x, \ldots, x)$ is always positive definite. Hence, the equal training symbols lead to a local minimum in our optimization problem.

To show this, we have:

$$\left.\frac{\partial^2\mathcal{J}}{\partial x_{i_m}\partial x_{i_q}^*}\right|_{(x,x,\ldots,x)} = \frac{L_h\sigma^2}{M\mathcal{E}^2}\sum_{l=1}^{\nu}e^{\jmath\frac{2\pi}{L_h}(m-q)((M-1)-\nu+l)} - \frac{\sigma^2}{2\mathcal{E}^2}, \quad q \neq m$$

and

$$\left.\frac{\partial^2\mathcal{J}}{\partial x_{i_m}\partial x_{i_m}^*}\right|_{(x,x,\ldots,x)} = \frac{3ML_h + 2L_h^2 - 2L_h - M}{M}\frac{\sigma^2}{2\mathcal{E}^2}$$

For an arbitrary vector $\boldsymbol{y}$ in the feasibility set of our optimization problem, we can write:

$$\begin{aligned}
\textstyle\sum_{k,m=0}^{L_h-1} \quad & y_k^* y_m \frac{\partial^2\mathcal{J}}{\partial x_{i_k}\partial x_{i_m}^*}|_{(x,x,\ldots,x)} \\
= \quad & \frac{L_h\sigma^2}{M\mathcal{E}^2}\sum_{k,m=0,k\neq m}^{L_h-1}y_k^* y_m\sum_{l=1}^{\nu}e^{\jmath\frac{2\pi}{L_h}(m-k)((M-1)-\nu+l)} \\
- \quad & \frac{\sigma^2}{2\mathcal{E}^2}\sum_{k,m=0,k\neq m}^{L_h-1}y_k^* y_m + \frac{3ML_h + 2L_h^2 - 2L_h - M}{M}\frac{\sigma^2}{2\mathcal{E}^2}\sum_{k=0}^{L_h-1}|y_k|^2 \\
= \quad & \frac{L_h\sigma^2}{M\mathcal{E}^2}\sum_{k,m=0}^{L_h-1}y_k^* y_m\sum_{l=1}^{\nu}e^{\jmath\frac{2\pi}{L_h}(m-k)((M-1)-\nu+l)} - \frac{L_h(L_h-1)}{M}\frac{\sigma^2}{\mathcal{E}^2}\sum_{k=0}^{L_h-1}|y_k|^2 \\
- \quad & \frac{\sigma^2}{2\mathcal{E}^2}\sum_{k,m=0,k\neq m}^{L_h-1}y_k^* y_m + \frac{3ML_h + 2L_h^2 - 2L_h - M}{M}\frac{\sigma^2}{2\mathcal{E}^2}\sum_{k=0}^{L_h-1}|y_k|^2
\end{aligned}$$

[7]In the complex field, we should consider the real part of the product of the Lagrange multiplier and the constraint. Nevertheless, in this problem, due to the symmetry of the constraint set, it can be proved that both approaches lead to the same result.

Furthermore,

$$\begin{aligned}\sum_{k,m=0,k\neq m}^{L_h-1} y_k^* y_m &= \Re\left\{\sum_{k,m=0,k\neq m}^{L_h-1} y_k^* y_m\right\} = \left|\sum_{k=0}^{L_h-1} y_k\right|^2 - \sum_{k=0}^{L_h-1}|y_k|^2 \\ &\overbrace{\leq}^{(a)} L_h \sum_{k=0}^{L_h-1}|y_k|^2 - \sum_{k=0}^{L_h-1}|y_k|^2 \\ &= (L_h-1)\sum_{k=0}^{L_h-1}|y_k|^2 \leq (L_h-1)\mathcal{E}\end{aligned}$$

where in $(a)$ we have used the Cauchy-Schwartz inequality, thus

$$-\frac{\sigma^2}{2\mathcal{E}^2}\sum_{k,m=0,k\neq m}^{L_h-1} y_k^* y_m \geq -\frac{(L_h-1)\sigma^2}{2\mathcal{E}^2}\sum_{k=0}^{L_h-1}|y_k|^2$$

We can write:

$$\frac{L_h\sigma^2}{M\mathcal{E}^2}\sum_{k,m=0}^{L_h-1} y_k^* y_m \sum_{l=1}^{\nu} e^{\jmath\frac{2\pi}{L_h}(m-k)((M-1)-\nu+l)} = \frac{L_h\sigma^2}{M\mathcal{E}^2}\sum_{l=1}^{\nu}\left|\sum_{m=0} y_m e^{\jmath\frac{2\pi}{L_h}m((M-1)-\nu+l)}\right|^2 \geq 0$$

The positivity of the Hessian is satisfied if

$$\left(\frac{3ML_h+2L_h^2-2L_h-M}{2M} - \frac{L_h(L_h-1)}{M} - \frac{L_h-1}{2}\right)\frac{\sigma^2}{\mathcal{E}^2} = L_h\frac{\sigma^2}{\mathcal{E}^2} > 0$$

which holds for any $L_h \geq 1$.

∎

The resulting (time-domain) MSE is:

$$\mathrm{MSE}_{L_h} = \frac{L_h\sigma^2}{\mathcal{E}}$$

## APPENDIX II

## CP-OFDM: THE CLASS OF EQUISPACED AND EQUAL PILOT TONES ACHIEVES THE GLOBAL MINIMUM MSE

In Appendix I, we proved that the class of equal training symbols is globally optimal for the CP-OFDM sparse preamble, when the pilot tones are equispaced. We now prove that this is also a globally optimal solution.

We will rule out the possibility that nonequispaced pilot tones can yield a lower MSE than the class of equispaced and equal training symbols. Consider again the MSE expression (38). We want to minimize this, subject to the constraint $\sum_{m=0}^{L_h-1}|x_{i_m}|^2 + \frac{1}{M}\boldsymbol{x}_{L_h}^H\boldsymbol{F}_{M\times\nu}\boldsymbol{F}_{M\times\nu}^H\boldsymbol{x}_{L_h} \leq \mathcal{E}$. It is obvious that since the CP is a wasted part of energy, the MSE would be minimized if somehow we were able to collect all the energy of the CP and put it in the useful part, namely the first sum of the constraint. To obtain the minimum MSE we can possibly imagine, we consider the relaxed problem:[8]

$$\min_{x_{i_m},\boldsymbol{F}_{L_h\times L_h}} \quad \mathrm{MSE}_{L_h} \tag{42}$$

$$\text{s.t.} \sum_{m=0}^{L_h-1}|x_{i_m}|^2 \leq \mathcal{E} \tag{43}$$

[8]This is a “genie-aided” problem, i.e., a problem that is unrealistic in practice and only a genie can help us to obtain, since it would lead to the minimum possible achievable MSE.

However, it is known [19], [3] that the optimal solution for this problem is the sparse preamble of $L_h$ equispaced and equipowered pilot tones. The minimum achievable MSE is $\frac{L_h \sigma^2}{\mathcal{E}}$. This is also achieved by a sparse preamble of equispaced and *equal* pilot tones in the previous appendix. Therefore, we only need to verify that this is the unique class of sparse preamble vectors that achieve this minimum MSE. This is equivalent to proving that the class of sparse preambles with equispaced and equal pilot tones is the only one that zeroes the CP energy.

The submatrix of $\boldsymbol{F}_{M\times\nu}\boldsymbol{F}_{M\times\nu}^H$ involved in the evaluation of $\boldsymbol{x}_{L_h}^H \boldsymbol{F}_{M\times\nu}\boldsymbol{F}_{M\times\nu}^H \boldsymbol{x}_{L_h}$ is given by $\left[\left(\boldsymbol{F}_{M\times\nu}\boldsymbol{F}_{M\times\nu}^H\right)_{i,j}\right]_{i,j\in\mathcal{I}_{L_h}}$ where $\mathcal{I}_{L_h}$ is now any set of the form $\{i_0 + kM/L_h |\, k = 0, 1, \ldots, L_h - 1\}$ with $i_0 = 0, 1, \ldots, M/L_h - 1$. This submatrix has a very special form:

*Lemma 2:* All diagonal entries of the submatrix $\left[\left(\boldsymbol{F}_{M\times\nu}\boldsymbol{F}_{M\times\nu}^H\right)_{i,j}\right]_{i,j\in\mathcal{I}_{L_h}}$ are equal to $\nu = L_h - 1$, while all its off-diagonal entries equal -1.

*Proof:* The general entry of the above submatrix is given by:

$$\left(\boldsymbol{F}_{M\times\nu}\boldsymbol{F}_{M\times\nu}^H\right)_{i,j} = \sum_{l=1}^{\nu} e^{\jmath\frac{2\pi}{L_h}(k_j-k_i)(M-1-\nu+l)} = \sum_{l=1}^{L_h-1} e^{\jmath\frac{2\pi}{L_h}(k_j-k_i)(M-L_h+l)}$$

where $i = i_0 + k_i \frac{M}{L_h}$ and similarly for $j$. Obviously, for $i = j$, $(\boldsymbol{F}_{M\times\nu}\boldsymbol{F}_{M\times\nu}^H)_{i,i} = L_h - 1$, $i = 0, 1, \ldots, L_h - 1$. If $i \neq j$, and setting $k = k_j - k_i$,

$$\left(\boldsymbol{F}_{M\times\nu}\boldsymbol{F}_{M\times\nu}^H\right)_{i,j} = \sum_{l=1}^{L_h-1} e^{\jmath\frac{2\pi}{L_h}k(M-L_h+l)} = \sum_{l=1}^{L_h-1} e^{\jmath\frac{2\pi}{L_h}kl}$$

with the assumptions made previously for $M, L_h$. But:

$$0 = \sum_{l=0}^{L_h-1} e^{\jmath\frac{2\pi}{L_h}kl} = 1 + \left(\boldsymbol{F}_{M\times\nu}\boldsymbol{F}_{M\times\nu}^H\right)_{i,j},$$

hence

$$\left(\boldsymbol{F}_{M\times\nu}\boldsymbol{F}_{M\times\nu}^H\right)_{i,j} = -1, \quad i \neq j$$

■

The question now concerns the type of vectors $\boldsymbol{x}_{L_h}$ that vanish the term $\boldsymbol{x}_{L_h}^H \boldsymbol{F}_{M\times\nu}\boldsymbol{F}_{M\times\nu}^H \boldsymbol{x}_{L_h} = \|\boldsymbol{F}_{M\times\nu}^H \boldsymbol{x}_{L_h}\|^2$ vanish. Suppose that there is such a sparse vector with equispaced and equipowered symbols $x_{i_m} = |x| e^{\jmath\theta_{i_m}}$. Then, there should hold $\boldsymbol{F}_{M\times\nu}^H \boldsymbol{x}_{L_h} = \mathbf{0}$, hence $\boldsymbol{F}_{M\times\nu}\boldsymbol{F}_{M\times\nu}^H \boldsymbol{x}_{L_h} = \mathbf{0}$. Consider, for example, the inner product of the first row of $\boldsymbol{F}_{M\times\nu}\boldsymbol{F}_{M\times\nu}^H$ with $\boldsymbol{x}_{L_h}$. Then, according to the previous lemma, there should hold $|x| \left(\nu e^{\jmath\theta_{i_0}} - \sum_{m=1}^{\nu} e^{\jmath\theta_{i_m}}\right) = 0$ or $\nu e^{\jmath\theta_{i_0}} = \sum_{m=1}^{\nu} e^{\jmath\theta_{i_m}}$. Taking the modulus in both sides, we should have $\nu = \left|\sum_{m=1}^{\nu} e^{\jmath\theta_{i_m}}\right|$. But this can only happen when all the exponentials in the sum are collinear and of the same direction in the complex plane, i.e., when all these exponentials are equal.

*Conclusion:* Among all sparse preamble vectors, it is those with equispaced and equal pilot symbols that yield the *globally* minimum MSE for CP-OFDM.

*Remark:* We can alternatively prove the statements of the last two Appendices without resorting to the Lagrange theory. Having defined the optimization problem (40)-(41), we can observe that the MSE achieved by the equispaced and equal pilot tones when they satisfy the constraint with equality is $L_h\sigma^2/\mathcal{E}$. Without proceeding with the proof of Proposition 1, we give the genie-aided problem (42)-(43) which achieves the minimum MSE for any sparse preamble. This minimum MSE is known to be $L_h\sigma^2/\mathcal{E}$ [19], [3]. Thus, we only need to verify that the class of equispaced and equal symbols is the unique MSE-optimal class for the sparse preamble design. This is shown as above.

## APPENDIX III
## CP-OFDM: OPTIMAL FULL PREAMBLE VECTORS WITH EQUIPOWERED PILOT TONES

The MSE expression (in the time domain) for the full preamble is $\mathrm{MSE}_M = \frac{1}{M^2}\mathrm{tr}\left[\boldsymbol{\mathcal{C}}_M\left(\boldsymbol{F}_{M\times L_h}\boldsymbol{F}^H_{M\times L_h}\right)\right]$, where $\boldsymbol{\mathcal{C}}_M$ is the estimation noise covariance matrix, which is diagonal, with diagonal entries of the form $\sigma^2/|x_m|^2$, $m = 0, 1, \ldots, M-1$. To obtain this expression, we have used the pseudo-inverse $\left(\boldsymbol{F}^H_{M\times L_h}\boldsymbol{F}_{M\times L_h}\right)^{-1}\boldsymbol{F}^H_{M\times L_h}$ to translate the CFR estimates to the time-domain and the fact that $\left(\boldsymbol{F}^H_{M\times L_h}\boldsymbol{F}_{M\times L_h}\right)^{-1} = (1/M)\boldsymbol{I}_{L_h}$. Clearly, here we do not face an optimal placement problem. Also, $\boldsymbol{F}_{M\times L_h}\boldsymbol{F}^H_{M\times L_h}$ is an $M\times M$ matrix with its diagonal elements all equal to $L_h$. Thus, the above MSE can be written as

$$\mathrm{MSE}_M = \frac{L_h}{M^2}\sum_{m=0}^{M-1}\frac{\sigma^2}{|x_m|^2} \tag{44}$$

Our problem then is to minimize this MSE subject to the constraint $\sum_{m=0}^{M-1}|x_m|^2 + \frac{1}{M}\boldsymbol{x}^H\boldsymbol{F}_{M\times\nu}\boldsymbol{F}^H_{M\times\nu}\boldsymbol{x} \le \mathcal{E}$. We already know that the training vector with all equal symbols is a global minimizer for that problem. In that case, $\mathrm{MSE}_M = \frac{L_h}{M^2}\frac{M\sigma^2}{|x|^2} = \frac{L_h}{M}\frac{\sigma^2}{\frac{\mathcal{E}}{M}} = \frac{L_h\sigma^2}{\mathcal{E}}$ and $\boldsymbol{x}^H\boldsymbol{F}_{M\times\nu}\boldsymbol{F}^H_{M\times\nu}\boldsymbol{x} = 0$. In the following, we show that there are also optimal full preamble vectors with simply equipowered, not necessarily equal symbols, and demonstrate ways of constructing them.

Obviously, if $\boldsymbol{x}^H\boldsymbol{F}_{M\times\nu}\boldsymbol{F}^H_{M\times\nu}\boldsymbol{x} = \|\boldsymbol{F}^H_{M\times\nu}\boldsymbol{x}\|^2 = 0$, then $\boldsymbol{x}$ must be spanned by the first $M-\nu$ columns of the $M\times M$ DFT matrix $\boldsymbol{\mathcal{F}}$. That is, it must be of the form $\boldsymbol{x} = \sum_{i=0}^{M-L_h}\alpha_i\boldsymbol{f}_i$, where $\boldsymbol{f}_i$, $i = 0,\ldots,M-L_h$, is the $i$th column of $\boldsymbol{\mathcal{F}}$. We observe that simply setting $\boldsymbol{x} = \sqrt{\mathcal{E}/M}\boldsymbol{f}_i$ for any $i = 0, 1, \ldots, M-1$ leads to equipowered (but unequal) symbols that minimize the MSE (cf. (44)). Also, due to the orthogonality of the DFT vectors and the fact that all of them have energy equal to $M$, we can see that the complex numbers $\alpha_i$, $i = 0, 1, \ldots, M-L_h$ should satisfy $\sum_{i=0}^{M-L_h}|\alpha_i|^2 = \frac{\mathcal{E}}{M}$.

We now give an algorithm for constructing an infinite number of such full preamble vectors, combining at most two of the first $M-\nu$ columns of $\boldsymbol{\mathcal{F}}$. Denote by $\boldsymbol{F}_{M\times(M-\nu)}$ the corresponding $M\times(M-\nu)$ matrix.

*Proposition 2:* We can find infinitely many $(M-\nu)$-tuples $\boldsymbol{\alpha} = \left[\begin{array}{cccc}\alpha_0 & \alpha_1 & \cdots & \alpha_{M-\nu-1}\end{array}\right]^T$ that satisfy $\|\boldsymbol{\alpha}\|^2 = \frac{\mathcal{E}}{M}$ and lead to $\boldsymbol{x} = \sum_{i=0}^{M-\nu-1}\alpha_i\boldsymbol{f}_i$, with $|x_i| = |x|$, $i = 0, 1, \ldots, M-1$, in the following two cases: First, only one of the $\alpha$'s, say $\alpha_m$, is nonzero, leading to a scaled version of the DFT column $\boldsymbol{f}_m$, and second, two of the $\alpha$'s are nonzero, say $\alpha_k, \alpha_m$, with phase differences $\pm\pi/2$ and $|k-m| = M/2$. The second case is only justified if $L_h < \frac{M}{2}$.

*Proof:* Clearly, the $\boldsymbol{\alpha}$'s we look for are such that $\boldsymbol{F}_{M\times(M-\nu)}\boldsymbol{\alpha} = \sqrt{\frac{\mathcal{E}}{M}}\boldsymbol{u}$, where $\boldsymbol{u}$ is any $M\times 1$ vector with unit modulus entries. For such a system of equations to be consistent, $\boldsymbol{u}$ should belong to the range space of $\boldsymbol{F}_{M\times(M-\nu)}$, i.e., $\boldsymbol{u} = \boldsymbol{F}_{M\times(M-\nu)}\boldsymbol{\gamma}$, for some complex vector $\boldsymbol{\gamma} = \begin{bmatrix} \gamma_0 & \gamma_1 & \cdots & \gamma_{M-\nu-1} \end{bmatrix}^T$. Then $\boldsymbol{\alpha} = \sqrt{\frac{\mathcal{E}}{M}}\boldsymbol{\gamma}$. Taking the squared norm of both sides of the last equation and using the constraint on the norm of $\boldsymbol{\alpha}$, we obtain

$$\sum_{i=0}^{M-\nu-1} |\gamma_i|^2 = 1$$

Moreover, taking the first entry of $\boldsymbol{u}$ and its modulus, we can write

$$\left|\sum_{i=0}^{M-\nu-1} \gamma_i\right| = 1$$

The last two equations are satisfiable in the following cases: First, if one of the $\gamma$'s is unit modulus, say $\gamma_m$, and the rest of them are zero. Then, $\boldsymbol{\alpha}$ will have only one nonzero entry, $\alpha_m$, with modulus $|\alpha_m| = \sqrt{\frac{\mathcal{E}}{M}}$. Alternatively, assume that only two of the $\gamma$'s are nonzero, say $\gamma_k, \gamma_m$, with the rest of them being zero. Then, if $\gamma_k$ has a modulus $\gamma$, i.e., $\gamma_k = \gamma e^{\jmath\theta}$, and $\gamma_m = \sqrt{1-\gamma^2}e^{\jmath(\theta\pm\frac{\pi}{2})}$, both equations are satisfied. In that case, there will only be two nonzero $\alpha$'s, namely $\alpha_k = \sqrt{\frac{\mathcal{E}}{M}}\gamma_k$ and $\alpha_m = \sqrt{\frac{\mathcal{E}}{M}}\gamma_m$. Hence, for the $r$th entry of $\boldsymbol{x}$, we will have $|x_r|^2 = \left|\alpha_k e^{-\jmath\frac{2\pi kr}{M}} + \alpha_m e^{-\jmath\frac{2\pi mr}{M}}\right|^2 = \frac{\mathcal{E}}{M} + 2\frac{\mathcal{E}}{M}\gamma\sqrt{1-\gamma^2}\Re\left\{e^{\jmath\left[\frac{2\pi(k-m)r}{M}\pm\frac{\pi}{2}\right]}\right\}$. This is obviously equal to $\mathcal{E}/M$ for $r = 0$, and also for $r \neq 0$ if $\Re\left\{e^{\jmath\left[\frac{2\pi(k-m)r}{M}\pm\frac{\pi}{2}\right]}\right\} = 0$, i.e., if $\frac{2\pi(k-m)r}{M} \pm \frac{\pi}{2} = \pm\frac{\pi}{2} \mod \pi$. This can be seen that it implies the requirement $|k-m| = \frac{M}{2}$. ∎

## APPENDIX IV
## OFDM/OQAM: OPTIMAL SPARSE AND FULL PREAMBLES

Define the vector of the nonzero SFB output samples for a sparse preamble input (cf. Section V-A):

$$\boldsymbol{s}_{\text{OQAM}}^{L_h} = \begin{bmatrix} \sum_{i\in\mathcal{I}_{L_h}} a_{i,0}g_{i,0}(0) & \sum_{i\in\mathcal{I}_{L_h}} a_{i,0}g_{i,0}(1) & \cdots & \sum_{i\in\mathcal{I}_{L_h}} a_{i,0}g_{i,0}(L_g-1) \end{bmatrix}^T$$

Clearly, for a sparse preamble, we have $M/L_h \geq 2$. We first show the following:

*Proposition 3:* If $M/L_h \geq 2$, then $\|\boldsymbol{s}_{\text{OQAM}}^{L_h}\|^2 = \sum_{i\in\mathcal{I}_{L_h}} a_{i,0}^2$, i.e., the energy transmitted for training is equal to the energy of the training vector at the AFB output of the associated ideal (channel- and noise-free) OFDM/OQAM system.

*Proof:*

$$\begin{aligned}
\|\boldsymbol{s}_{\text{OQAM}}^{L_h}\|^2 &= \sum_{l=0}^{L_g-1}\left|\sum_{i\in\mathcal{I}_{L_h}} a_{i,0}g_{i,0}(l)\right|^2 = \sum_{m=0}^{L_h-1} a_{i_m,0}\sum_{k=0}^{L_h-1} a_{i_k,0}\sum_{l=0}^{L_g-1} g_{i_m,0}(l)g_{i_k,0}^*(l) \\
&= \sum_{m=0}^{L_h-1} a_{i_m,0}^2\sum_{l=0}^{L_g-1}|g_{i_m,0}(l)|^2 + \sum_{m,k=0,m\neq k}^{L_h-1} a_{i_m,0}a_{i_k,0}\sum_{l=0}^{L_g-1} g_{i_m,0}(l)g_{i_k,0}^*(l)
\end{aligned}$$

Obviously,

$$\sum_{l=0}^{L_g-1}|g_{i_m,0}(l)|^2 = \sum_{l=0}^{L_g-1} g(l)^2 = 1$$

and

$$\sum_{l=0}^{L_g-1} g_{i_m,0}(l)g_{i_k,0}^*(l) = 0, \quad m,k = 0,1,\ldots,L_h-1,\ m\neq k$$

due to our assumption that $M/L_h \geq 2$ and a good frequency localization of $g$. Therefore, $\|\boldsymbol{s}_{\mathrm{OQAM}}^{L_h}\|^2 = \sum_{m=0}^{L_h-1} a_{i_m,0}^2$, $i_m \in \mathcal{I}_{L_h}$. ■

The (time domain) MSE expression for the sparse preamble with $L_h$ nonzero pilot tones in the OFDM/OQAM system is the same as in the CP-OFDM system, i.e., $\mathrm{MSE}_{L_h} = \mathrm{tr}\left[\boldsymbol{\mathcal{C}}_{L_h}\left(\boldsymbol{F}_{L_h\times L_h}\boldsymbol{F}_{L_h\times L_h}^H\right)^{-1}\right]$. Our optimization problem can therefore be stated as follows:

$$\min_{a_{i_m}, i_m \in \mathcal{I}_{L_h}} \mathrm{MSE}_{L_h} \tag{45}$$

$$\text{s.t.} \sum_{m=0}^{L_h-1} a_{i_m}^2 \leq \mathcal{E} \tag{46}$$

where we have suppressed the temporal index $0$. But the solution to this problem is known. It is the class of equipowered and equispaced pilot tones [19], [3].

For the full preamble, it is easy to show that the transmit training energy is *not* equal to the energy of the training vector at the AFB output. Using our assumption on the time-frequency localization of the prototype function, the training energy constraint can be written as:

$$\sum_{m=0}^{M-1}\left(|x_m|^2 + \beta x_m x_{m-1}^* + \beta x_m x_{m+1}^*\right) \leq \mathcal{E},$$

where $x_m = a_{m,0}e^{\jmath\varphi_{m,0}}$, and therefore the optimization problem can be stated as:

$$\min_{x_m, m=0,1,\ldots,M-1} \mathrm{MSE}_M = \frac{1}{M}\sum_{m=0}^{M-1}\frac{\sigma^2}{|x_m + \beta x_{m-1} + \beta x_{m+1}|^2} \tag{47}$$

$$\text{s.t.} \quad \sum_{m=0}^{M-1}\left(|x_m|^2 + \beta x_m x_{m-1}^* + \beta x_m x_{m+1}^*\right) \leq \mathcal{E} \tag{48}$$

Using similar steps as in the proof of Proposition 1, we can easily show that the full preamble vector with all equal symbols is a minimizer of (47), (48) and the minimum achievable MSE is $M\sigma^2/\left[\mathcal{E}(1+2\beta)^2\right]$. Furthermore, we can show that this preamble is a global minimizer of the last optimization problem, but with a constraint on the training energy at the SFB input, i.e., with a constraint of the form:

$$\sum_{m=0}^{M-1}|x_m|^2 \leq \mathcal{E}$$

*Proposition 4:* A global minimizer of the problem:

$$\min_{x_m, m=0,1,\ldots,M-1} \mathrm{MSE}_M = \frac{1}{M}\sum_{m=0}^{M-1}\frac{\sigma^2}{|x_m + \beta x_{m-1} + \beta x_{m+1}|^2} \tag{49}$$

$$\text{s.t.} \quad \sum_{m=0}^{M-1}|x_m|^2 \leq \mathcal{E} \tag{50}$$

is the full preamble with *equal* symbols.

*Proof:* To prove the statement of this proposition we can initially show that the equal symbols is a local minimizer of our optimization problem via Lagrange theory and then verify that the equal symbols lead to the minimum possible MSE. However, if we show that the equal symbols achieve the lowest MSE, the step associated with the Lagrange theory in unnecessary[9].

We can divide the class of symbols into the following subclasses:

1a) Equal symbols, i.e., symbols of the same magnitude and phase[10].

1b) Symbols of equal modulus and different phases (at least one symbol with different phase from the rest of the symbols in the preamble vector).

2a) Symbols of different modulus but of the same phase (at least one symbol with different modulus than the rest of the symbols in the preamble vector).

2b) Symbols of different modulus and phase (at least a symbol different from the others in the preamble vector).

The subclasses 1a), 1b) subdivide the general class of *equipowered* symbols and subclasses 2a), 2b) the general class of *nonequipowered* symbols. First, we will show that between subclasses 1a) and 1b), 1a) leads to an equal or lower MSE than that of 1b). The same holds for the subclass 2a), when it is compared with 2b).

*Comparison of 1a), 1b) subclasses*: We consider a vector of equal symbols $x_0 = x_1 = \cdots = x_{M-1} = x = |x|e^{\jmath\phi}$ and an arbitrary vector of the form $x_0 = |x|e^{\jmath\phi_0}, x_1 = |x|e^{\jmath\phi_1}, \ldots, x_{M-1} = |x|e^{\jmath\phi_{M-1}}$. For the case of equal symbols, the arbitrary term $\frac{\sigma^2}{|x_m+\beta x_{m-1}+\beta x_{m+1}|^2}$ takes the value $\frac{\sigma^2}{|x|^2(1+2\beta)^2}$ and if we assume that the constraint is satisfied with the equality, then this value becomes $\frac{M\sigma^2}{\mathcal{E}(1+2\beta)^2}$. In the case of unequal symbols, the modulus is again $\mathcal{E}/M$ since the constraint is phase blind, thus the maximum value that can be taken by any such term is $\frac{M\sigma^2}{\mathcal{E}(1+2\beta)^2}$. It is obvious that in any way the phases $\phi_0, \phi_1, \ldots, \phi_{M-1}$ are chosen, there is at least one term $\frac{\sigma^2}{|x_m+\beta x_{m-1}+\beta x_{m+1}|^2}$ for the unequal symbols that is greater than or equal to $\frac{M\sigma^2}{\mathcal{E}(1+2\beta)^2}$ (essentially, we can not achieve in all cases triplets of numbers $x_{m-1}, x_m, x_{m+1}$ that are collinear and of the same directionality in the complex plane). Therefore:

$$\mathrm{MSE}^{1a} \le \mathrm{MSE}^{1b}$$

In the same way:

$$\mathrm{MSE}^{2a} \le \mathrm{MSE}^{2b}$$

To finish this proof, we have to show that $\mathrm{MSE}^{1a} \le \mathrm{MSE}^{2a}$. For the subclass 1a) we can write:

$$\begin{aligned}
\mathrm{MSE}^{1a} &= \frac{1}{M}\sum_{m=0}^{M-1}\frac{\sigma^2}{|x_m+\beta x_{m-1}+\beta x_{m+1}|^2} = \frac{M\sigma^2}{\sum_{m=0}^{M-1}|x_m+\beta x_{m-1}+\beta x_{m+1}|^2} \\
&= \frac{M\sigma^2}{\mathcal{E}(1+2\beta)^2}
\end{aligned}$$

using the Arithmetic-Geometric-Harmonic (AGH) mean inequality. For the subclass 2a), we have:

$$\mathrm{MSE}^{2a} = \frac{1}{M}\sum_{m=0}^{M-1}\frac{\sigma^2}{|x_m+\beta x_{m-1}+\beta x_{m+1}|^2} \ge \frac{M\sigma^2}{\sum_{m=0}^{M-1}|x_m+\beta x_{m-1}+\beta x_{m+1}|^2}$$

[9]Note that this holds also for the proof of Proposition 1. I.e., we can alternatively combine the results in the first two appendices, discarding the Lagrange theory step and simply verifying that the equal and equispaced symbols achieve the lowest MSE, which is equal to $L_h\sigma^2/\mathcal{E}$.

[10]We refer to the interval $[0, 2\pi)$ for the phases, since for a phase $\phi_0 \in [0, 2\pi]$, the phase $\phi_0 \pm 2k\pi, k \in \mathbb{Z}$ leads to the same phasor $e^{\jmath\phi_0}$.

If we show that:

$$\sum_{m=0}^{M-1} |x_m + \beta x_{m-1} + \beta x_{m+1}|^2 \leq \mathcal{E}(1+2\beta)^2$$

or

$$\sum_{m=0}^{M-1} (\zeta_m + \beta \zeta_{m-1} + \beta \zeta_{m+1})^2 \leq \mathcal{E}(1+2\beta)^2$$

we will be done. Here, $\zeta_m = |x_m|$ and the last equation is obtained due to the equal phases of the symbols.

For the last expression, we have:

$$\begin{aligned} \textstyle\sum_{m=0}^{M-1} \quad & (\zeta_m + \beta \zeta_{m-1} + \beta \zeta_{m+1})^2 = \sum_{m=0}^{M-1} \zeta_m^2 + \beta^2 \sum_{m=0}^{M-1} \zeta_{m-1}^2 + \beta^2 \sum_{m=0}^{M-1} \zeta_{m+1}^2 \\ + \quad & 2\beta \sum_{m=0}^{M-1} \zeta_m \zeta_{m-1} + 2\beta \sum_{m=0}^{M-1} \zeta_m \zeta_{m+1} + 2\beta^2 \sum_{m=0}^{M-1} \zeta_{m-1} \zeta_{m+1} \end{aligned}$$

We can consider that the index $m-1$ for $m=0$ equals $M-1$ and correspondingly, the index $m+1$ for $m = M-1$ equals $0$ due to the periodicity of the discrete time spectrum. Therefore:

$$\sum_{m=0}^{M-1} \zeta_m^2 = \sum_{m=0}^{M-1} \zeta_{m-1}^2 = \sum_{m=0}^{M-1} \zeta_{m+1}^2 = \mathcal{E}$$

For the same reason:

$$\sum_{m=0}^{M-1} \zeta_m \zeta_{m-1} = \sum_{m=0}^{M-1} \zeta_m \zeta_{m+1}$$

We set $\zeta_{m-1} = \xi_m$ in the last equation. We place the numbers $\zeta_m$ and $\xi_m$ to the main diagonals of two diagonal matrices $\boldsymbol{Z}$ and $\boldsymbol{\Xi}$, respectively. The following lemma [8, p. 183] will prove useful:

*Lemma 3:* Consider any two matrices $\boldsymbol{A}, \boldsymbol{B} \in \mathbb{C}^{m \times n}$ and their singular values, $\sigma_i(\boldsymbol{A}), \sigma_i(\boldsymbol{B})$, $i = 1, 2, \ldots, q$, with $q = \min\{m, n\}$, arranged in a descending order. Then:

(a) The following inequality holds: $|\mathrm{tr}(\boldsymbol{A}^H \boldsymbol{B})| \leq \sum_{i=1}^q \sigma_i(\boldsymbol{A}) \sigma_i(\boldsymbol{B})$.

(b) There exist unitary matrices $\boldsymbol{P}_1$ and $\boldsymbol{P}_2$ such that $\max\{|\mathrm{tr}(\boldsymbol{P}_1 \boldsymbol{A}^H \boldsymbol{P}_2 \boldsymbol{B})| : \boldsymbol{P}_1 \in \mathbb{C}^{n \times n}, \boldsymbol{P}_2 \in \mathbb{C}^{m \times m} \text{ are unitary}\} = \sum_{i=1}^q \sigma_i(\boldsymbol{A}) \sigma_i(\boldsymbol{B})$.

Using the last lemma, we write:

$$\mathrm{tr}\,(\boldsymbol{Z}\boldsymbol{\Xi}) = \sum_{m=0}^{M-1} \zeta_m \zeta_{m-1} \leq \sum_{m=0}^{M-1} \zeta_m^2 = \mathcal{E}$$

Using the same trick, we can show that:

$$\sum_{m=0}^{M-1} \zeta_{m+1} \zeta_{m-1} \leq \sum_{m=0}^{M-1} \zeta_m^2 = \mathcal{E}$$

Therefore,

$$\begin{aligned} \sum_{m=0}^{M-1} (\zeta_m + \beta \zeta_{m-1} + \beta \zeta_{m+1})^2 & \leq \mathcal{E} + \beta^2 \mathcal{E} + \beta^2 \mathcal{E} + 4\beta \mathcal{E} + 2\beta^2 \mathcal{E} \\ & = \mathcal{E}(1+2\beta)^2 \end{aligned}$$

and our proposition is proved. ■

# APPENDIX V

## OFDM/OQAM: MSE WITH POSTPROCESSING OF THE FULL-PREAMBLE-BASED ESTIMATES

To derive the MSE expression, we first have to determine the covariance of the noise at the AFB output. The $m$th noise component at the output of the AFB of the OFDM/OQAM system corresponding to the preamble vector is given by $\eta_{m,0} = \sum_{l=0}^{L_g-1} w(l)g_{m,0}^*(l)$, where $w(l)$ is the noise at the receiver front-end, assumed additive and white with zero mean and variance $\sigma^2$. We then have:

$$E\left(\eta_{m,0}\eta_{k,0}^*\right) = \sum_l \sum_r E\left[w(l)w^*(r)\right] g_{m,0}^*(l)g_{k,0}(r) = \sigma^2 \sum_l \sum_r \delta_{l,r} g_{m,0}^*(l)g_{k,0}(r) = \sigma^2 \sum_l g_{m,0}^*(l)g_{k,0}(l)$$

For $m = k$, this is equal to $\sigma^2$. According to our assumptions on the good localization of the pulse in frequency, and using the fact that the phases, $\varphi_{m,0}$, of the symbols are equal, we can see that the cross-correlation term above takes the value $\sigma^2\beta$ for $l = m-1, m+1$ and is zero elsewhere. Therefore, the noise covariance matrix at the AFB output can be expressed as:[11]

$$\boldsymbol{\mathcal{C}}_M = \sigma^2 \underbrace{\begin{bmatrix} 1 & \beta & 0 & \cdots & \pm\beta \\ \beta & 1 & \beta & \cdots & 0 \\ \vdots & \vdots & \ddots & \ddots & \vdots \\ \pm\beta & 0 & \cdots & \beta & 1 \end{bmatrix}}_{\boldsymbol{B}}$$

When computing the initial CFR estimates as in (10), the noise covariance matrix becomes $\boldsymbol{\mathcal{C}}'_M = \frac{\sigma^2}{\mathcal{E}_x^{\mathrm{OQAM}}(1+2\beta)^2}\boldsymbol{B}$, and hence the corresponding MSE is $M\sigma^2/\left[\mathcal{E}_x^{\mathrm{OQAM}}(1+2\beta)^2\right]$ as in (18). After the processing described in Section V-C, we come up with the final CFR estimates, with MSE given by:

$$\begin{aligned} \mathrm{MSE} &= \frac{\sigma^2}{\mathcal{E}_x^{\mathrm{OQAM}}(1+2\beta)^2}\mathrm{tr}\left[\boldsymbol{F}_{M\times L_h}\left(\boldsymbol{F}_{M\times L_h}^H\boldsymbol{F}_{M\times L_h}\right)^{-1}\boldsymbol{F}_{M\times L_h}^H\boldsymbol{B}\boldsymbol{F}_{M\times L_h}\left(\boldsymbol{F}_{M\times L_h}^H\boldsymbol{F}_{M\times L_h}\right)^{-1}\boldsymbol{F}_{M\times L_h}^H\right] \\ &= \frac{\sigma^2}{\mathcal{E}_x^{\mathrm{OQAM}}(1+2\beta)^2}\frac{1}{M}\mathrm{tr}\left(\boldsymbol{B}\boldsymbol{F}_{M\times L_h}\boldsymbol{F}_{M\times L_h}^H\right) \end{aligned}$$

$\boldsymbol{F}_{M\times L_h}\boldsymbol{F}_{M\times L_h}^H$ has all its main diagonal entries equal to $L_h$. For large $M$, it is easy to see that its entries immediately above and below its main diagonal can also be well approximated by $L_h$. This leads to:

$$\mathrm{MSE} \approx \frac{L_h\sigma^2}{\mathcal{E}_x^{\mathrm{OQAM}}(1+2\beta)}$$

[11] The $\pm$'s depend on the choice of the prototype filter.